\documentclass[12pt,preprint]{aastex}

\def\lesssim{\mathrel{\hbox{\rlap{\hbox{\lower4pt\hbox{$\sim$}}}\hbox{$<$}}}}
\def\gtrsim{\mathrel{\hbox{\rlap{\hbox{\lower4pt\hbox{$\sim$}}}\hbox{$>$}}}}
\providecommand{\etal}{et~al.}

\begin{document}
\title{The First {\em Swift} Ultra-Violet/Optical Telescope GRB Afterglow 
Catalog}
\author{P.~W.~A. Roming\altaffilmark{1}, T.~S. Koch\altaffilmark{1}, S.~R. Oates\altaffilmark{2}, 
B.~L. Porterfield\altaffilmark{3}, D.~E. Vanden Berk\altaffilmark{1}, 
P.~T. Boyd\altaffilmark{4}, S.~T. Holland\altaffilmark{4,5,6}, E.~A. Hoversten\altaffilmark{1},
S. Immler\altaffilmark{4,5}, 
F.~E. Marshall\altaffilmark{4}, M.~J. Page\altaffilmark{2}, J.~L. Racusin\altaffilmark{1}, D.~P. Schneider\altaffilmark{1},
A.~A. Breeveld\altaffilmark{2}, P.~J. Brown\altaffilmark{1}, M.~M. Chester\altaffilmark{1},
A. Cucchiara\altaffilmark{1}, M. De~Pasquale\altaffilmark{2}, C. Gronwall\altaffilmark{1},
S.~D. Hunsberger\altaffilmark{1}, N.~P.~M. Kuin\altaffilmark{2}, 
W.~B. Landsman\altaffilmark{4}, P. Schady\altaffilmark{2},
M. Still\altaffilmark{2}} 

\altaffiltext{1}{Department of Astronomy \& Astrophysics,
Penn State University, 525 Davey Lab, University Park, PA
16802, USA; Corresponding author's e-mail: roming@astro.psu.edu}

\altaffiltext{2}{Mullard Space Science Laboratory, University College London, 
Holmbury St. Mary, Dorking, Surrey RH5 6NT, UK}

\altaffiltext{3}{State College Area High School, 653 Westerly Parkway, State 
College, PA 16801, USA}

\altaffiltext{4}{NASA/Goddard Space Flight Center, Greenbelt, MD 20771, USA}

\altaffiltext{5}{Universities Space Research Association, 10227 Wincopin Circle, 
Suite 500, Columbia, MD 21044, USA}

\altaffiltext{6}{Centre for Space Research in Space Science and Technology,
Code 660, NASA/GSFC, 8800 Greenbelt Rd, Greenbelt, MD 20771, USA}

\begin{abstract}
We present the first {\em Swift} Ultra-Violet/Optical Telescope (UVOT)
gamma-ray burst (GRB) afterglow catalog. The catalog contains data from
over $64,000$ independent UVOT image observations of 229 GRBs first
detected by {\em Swift}, the {\em High Energy Transient Explorer~2}
(HETE2), the {\em INTErnational Gamma-Ray Astrophysics Laboratory}
(INTEGRAL), and the Interplanetary Network (IPN). The catalog
covers GRBs occurring during the period from 2005 Jan 17 to 2007 Jun 16 and 
includes $\sim86\%$ of the bursts detected by the {\em Swift} 
Burst Alert Telescope (BAT). The catalog provides
detailed burst positional, temporal, and photometric information extracted
from each of the UVOT images. Positions for bursts detected at the 
$3\sigma$-level are
provided with a nominal accuracy, relative to the USNO-B1 catalog, of
$\sim0\farcs25$. Photometry for each burst is given in three UV bands,
three optical bands, and a `$white$' or open filter. Upper limits for
magnitudes are reported for sources detected below $3\sigma$. General properties
of the burst sample and light curves, including the filter-dependent
temporal slopes, are also provided. The majority of the UVOT
light curves, for bursts detected at the 
$3\sigma$-level, can be fit by a single power-law,
with a median temporal slope ($\alpha$) of 0.96, beginning several hundred
seconds after the burst trigger and ending at $\sim1\times10^5 {\rm ~s}$. 
The median UVOT $v$-band ($\sim5500 {\rm ~\AA}$) magnitude 
at $2000 {\rm ~s}$ for a sample of ``well" detected bursts 
is 18.02. The UVOT flux interpolated
to $2000 {\rm ~s}$ after the burst, shows relatively strong correlations with
both the prompt {\em Swift} BAT fluence, and the {\em Swift} X-ray
flux at $11 {\rm ~hours}$ after the trigger. 
\end{abstract}


\section {Introduction}
The Ultraviolet/Optical Telescope \citep[UVOT;][]{RPWA2005} on board
the {\em Swift} \citep{GN2004} observatory, provides rapid follow-up
observations of GRB afterglows at UV and optical wavelengths. Typical
times from a {\em Swift} Burst Alert Telescope \citep[BAT;][]{BS2005}
GRB trigger to first UVOT observation range from 40 to 200 seconds.
The UVOT usually provides the first optical, and almost always the only
UV observations, of GRB afterglows. For detected GRB afterglows, the UVOT
provides rapidly available sub-arcsecond locations, as well as high time
resolution light curves in up to seven broad photometric bands. The relatively
deep UVOT detection limits at early times also place constraints on the
``dark burst'' phenomenon \citep[cf.][]{RPWA06b}. In addition, the UVOT observations are made
simultaneously with X-ray observations with the coaligned {\em Swift}
X-Ray Telescope \citep[XRT;][]{BDN2005}, allowing direct comparison of
burst properties in X-ray and UV/optical light.

This paper describes the first {\em Swift} UVOT GRB afterglow catalog,
containing information on bursts observed during the first 2.5 years
of UVOT operation (2005-2007; corresponding to the instrument
turn-on until the end of the 1st BAT GRB catalog \citep{ST2007}). 
The catalog includes 205 bursts first detected by the
{\em Swift} BAT, and 24 bursts first detected by other satellites
or networks: the {\em High Energy Transient Explorer~2}
\citep[HETE2;][]{RG1997}, the {\em INTErnational Gamma-Ray Astrophysics
Laboratory} (INTEGRAL), and
the Interplanetary Network \citep[IPN;][]{HK05}.  Information is provided
for over $64,000$ individual images of GRB fields taken by the UVOT.

In Section 2, we present the observations made by the UVOT. In 
Section 3, we describe the construction of the databases and
the catalog. In Section 4, we describe the format of the databases
and catalog. In Section 5, we provide a summary
of the catalog. In Section 6 we summarize future work to be 
accomplished. The databases and catalog are provided in electronic format with
this paper and are also available at the {\em Swift}
website\footnote{http://swift.gsfc.nasa.gov/docs/swift/results/uvot\_grbcat/}.

\section{Observations}
The UVOT utilizes seven broadband filters during the observation of
GRBs. The characteristics of the filters --- central wavelength
($\lambda_c$), FWHM, zero points (the magnitudes at which the
detector registers $1 {\rm ~count\,s^{-1}}$; $m_z$), and the flux
conversion factors ($f_{\lambda}$) --- can be found in Table~\ref{tab5}
\citep{PTS2007,RPWA2005}. The flux density conversion factors are
calculated based on model GRB power law spectra with a redshift ranging
from $0.3 < z < 1.0$ \citep{PTS2007}\footnote{The most recent calibration
data are available from the {\em Swift} calibration database at
http://swift.gsfc.nasa.gov/docs/heasarc/caldb/swift/}. The nominal
image scale for UVOT images is $0\farcs502 {\rm ~pixel^{-1}}$ (unbinned). 
UVOT data is collected in one of
two modes: event (or photon counting) and image. Event mode captures the
time of the arriving photon as well as the celestial coordinates. The temporal
resolution in this mode is $\sim11\,{\rm ~ms}$. In image mode, photons
are counted during the exposure and the position is recorded, 
but no timing information is stored
except for the start and stop times of the exposure.
Because the spacecraft has limited data storage capabilities,
most UVOT observations are performed in image mode since the 
telemetry rate is significantly lower than event mode observations.

Since the launch of {\em Swift}, the automated observing sequence
of the UVOT has been changed a few times in order to optimize observations of
GRBs. The automated sequence is a set of variables which includes,
but is not limited to, the filters, modes, and exposure times. The
basic automated sequence design consists of finding charts and a series
of short, medium, and long exposures in various filters. The finding
charts are typically taken in both event and image mode simultaneously,
in both the $white$ and $v$ filters, and have exposure times ranging
from $100-400\,{\rm ~s}$. A subset of these finding charts are
immediately telemetered to ground-based telescopes to aid in
localizing the GRB\footnote{A
discussion of the finding chart and simultaneous observations
in event and image mode is beyond the scope of this paper. The reader
is referred to \citet{RPWA2005}. For observations between 2006 Jan 10
to 2006 Feb 13, and 2006 Mar 15 to the present, a second set of
finding charts was included in the sequence. The second set
of finding chart exposures are taken in the same way as the first set,
except that exposures in the $white$ filter are taken in image mode only.}.
After completion of the $white$ and $v$ finding charts, a series of 
short exposures is typically taken in event mode, in all
seven broadband filters, and has exposure times ranging from 
$10-50\,{\rm ~s}$. A series of medium exposures is then taken in image mode,
in all seven broadband filters, and has exposure times ranging from
$100-200\,{\rm ~s}$. Finally, a series of long exposures is taken in
image mode, in all seven broadband filters,
and has typical exposure times of $900\,{\rm ~s}$.\footnote{Between 
2005 Jan 17 to 2006 Jan 9 and between
2006 Feb 24 to 2006 Mar 14, exposures taken in uvw2, uvm2,
and uvw1 were taken in event mode.} In all cases, exposures
can be cut short due to observing constraints.

This catalog covers UVOT observations of 229 GRB afterglows
from 2005 Jan 17 to 2007 Jun 16. It includes bursts detected
by {\em Swift} BAT, HETE2, INTEGRAL, IPN and observed by UVOT.
A total of 211 BAT-detected bursts were observed by the UVOT (after
instrument turn on) representing 93\% of the BAT sample. Those that were
not observed by the UVOT were either too close in angular distance to
a bright ($\sim 6 {\rm ~mag}$) source, or occurred during
UVOT engineering observations. Not included in the catalog are nine
bursts first detected by BAT and INTEGRAL and observed by UVOT but with no
afterglow position reported by the XRT or ground based observers (see
Table~\ref{tab6}). Inspection of the UVOT images reveals no obvious 
afterglows for these bursts.

Hereafter, we adopt the notation $F(\nu ,t) \propto t^{-\alpha} \nu^{-\beta}$ 
for the afterglow flux density as a function of time, where
$\nu$ is the frequency of the observed flux density, $t$ is the time post
trigger, $\beta$ is the spectral index which is related to the
photon index $\Gamma$ ($\beta = \Gamma - 1$) , and $\alpha$ is
the temporal decay slope. 

\section{Construction of the Databases and Catalog}
To provide a baseline for understanding the work
described below, we define three words in the 
context of this catalog: database,
catalog, and photometry pipeline. The database is the 
repository for all UVOT GRB data processed by the 
photometry pipeline. The catalog is the compilation of 
the top-level data 
from the UVOT database, as well as other sources (i.e.
the BAT catalog \citep{ST2007}, the Gamma-ray burst
Coordinate Network \citep[GCN;][]{BS1995, BS1998} Circulars, etc.) that provides
the primary characteristics of each burst. The photometry 
pipeline is the script that 
combines the required FTOOLs (with {\tt uvotsource}
performing the photometry) 
to produce the database and catalog. Below we describe
the construction of the image and event databases, the
catalog, and the requisite quality checks.

\subsection{Image Database Construction}
The UVOT GRB photometric image database is a collection of raw 
photometric measurements of UVOT images. The first step in 
constructing the database was to build an archive of UVOT GRB 
images. To ensure that all of our images and exposure maps 
benefitted from consistent and up-to-date calibrations and 
processing, the {\em Swift} Data Center reprocessed the entire 
UVOT GRB image archive\footnote{Version 
HEA\_06DEC2006\_V6.1.1\_SWIFT\_REL2.6(BLD20)PATCHED3\_14MAR2007 
was used.}. For example, images taken early in the 
mission did not benefit from MOD-8 pattern noise correction. 
All reprocessed images are now MOD-8 corrected. A number of 
images in our archive did not have a fine aspect correction 
applied; many of these images were recovered by running 
{\tt uvotskycorr}. Lastly, we corrected a number of images 
which had improper OBJECT keywords. 

\begin{figure}
\epsscale{0.9}
\plotone{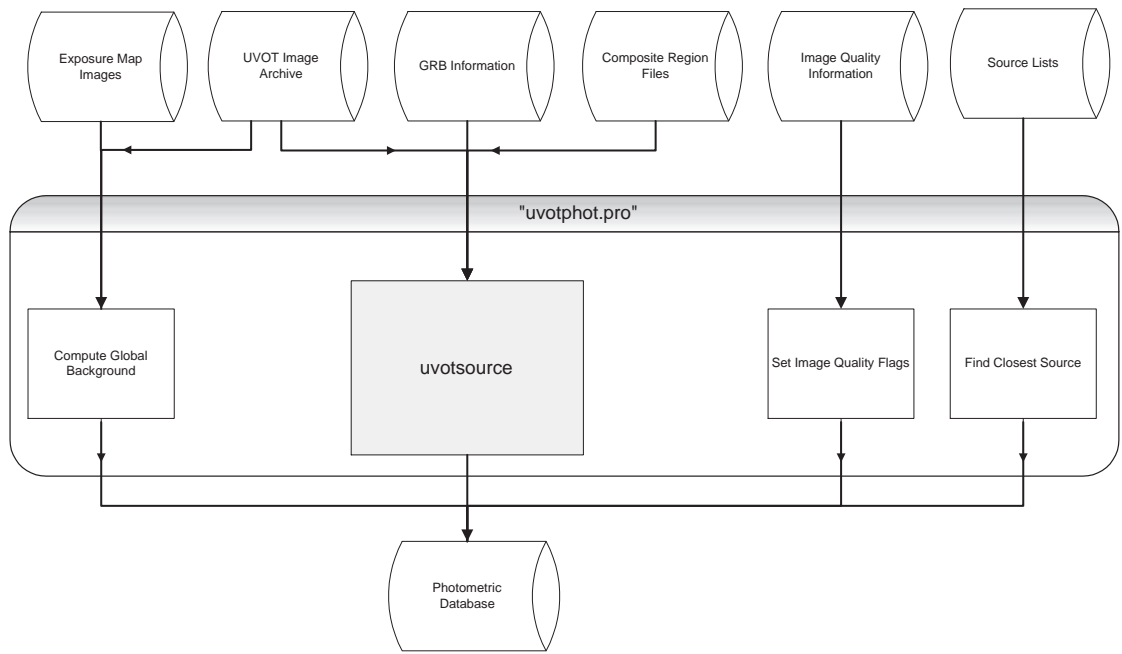}
\caption{Flowchart of the photometric pipeline software
used to create the database.}
\label{fig-flowchart}   
\end{figure}

We developed an IDL based image processing pipeline to 
perform aperture photometry on our archive of 
re-processed sky images. Figure~\ref{fig-flowchart} is a diagram 
of the UVOT {\tt uvotphot} photometric pipeline software 
(Version 1.0). The heart of {\tt uvotphot} is the {\em Swift} 
UVOT tool {\tt uvotsource}. The pipeline used HEADAS Version 6.4 
and the 2007 November UVOT CALDB. 
Photometry was performed on individual sky images using 
the curves of growth from the {\em Swift} CALDB for the aperture correction
model. Upper limits were reported for sources $< 3\sigma$.

Figure~\ref{fig-flowchart} shows the additional inputs to the uvotphot pipeline 
software. The GRB information file contains the best reported 
source position and error estimate for each burst. In a small 
number of cases, we refined the positions to better center 
bursts in our source apertures. References for the best 
reported burst positions can be found in Table~\ref{tab4}.
Source region files specify a simple circular inclusion 
region of a given aperture size. Aperture photometry using 
a $3.0\arcsec$ radius aperture was performed for 
Version 1.0 of the database. Since a $5.0\arcsec$ radius aperture 
(containing $85.8\pm3.8\%$ of the PSF) was used for calibrating
the UVOT, an aperture correction is applied to the data \citep{PTS2007}.

Background regions contain inclusion regions for background 
estimation and exclusion regions to mask out sources and/or 
features in each field. In most cases we use a standard 
annular inclusion region of inner and outer radii of $27\farcs5$ 
and $35\farcs0$
around each burst. To decrease the number of background 
region files required by our pipeline, we constructed 
composite region files to mask out sources and features in 
all bands. A small number of fields required non-standard 
region files. The background is calculated by taking the
average background of all the pixels in the net background
region. The region files (and postage stamp images) 
used in our processing can be found at the {\em Swift} website. 

Sources within $15\arcsec$ of each burst, which may 
contaminate photometry in our source apertures, are listed 
in the {\tt UvotSourceTable.fits} product. For each source 
we record its position and the measured magnitude in each 
filter. Source contamination is evidenced as a non-zero 
offset in the light curve data products.

UVOT images which suffer from other sources of contamination 
or degradation (not including nearby sources) have been 
identified. Image quality information is recorded as a 
series of ``flags" which correspond to 1) bursts embedded in 
a large halo structure from nearby bright stars, 2) images 
where the burst was near the edge of the FoV, 3) images with 
charge trails in the source aperture, 4) images with 
diffraction spikes in the source aperture, 5) images which 
do not have fine aspect corrections, and 6) bursts 
embedded in crowded fields. Flags set to true (``T") indicate 
images which are of poor quality. Images considered to be of low quality 
and have at least one quality flag raised also have the Quality Flag 
set to true (``T").

For images that were generated from both image and event mode
simultaneously (usually the finding charts are taken in this
dual mode), the event mode data are excluded from the image
database. A description of the event mode data can be found in 
Section 3.2 below.

\subsection{Event Database Construction}
For each GRB, the first orbit $v$ and $white$ event 
lists were obtained from the the Swift Data Center. 
Like the sky images, the event lists benefited from 
reprocessing using UVOT(20071106) and MIS(20080326) 
calibration and 
processing pipeline. The $v$ and $white$ are the 
only filters for which event lists have been 
included in the catalog because these filters are used
for the finding chart exposures, which 
are expected to show variability on the shortest 
timescales. 
 
The event processing pipeline is based on PERL and 
implements the {\it Swift} software and calibration 
found in the HEADAS 6.3.2 release. The pipeline 
refines the aspect of the event lists and then 
extracts the photometry. During the slew, 
immediately after the BAT trigger, the star tracker 
reports step-like changes in position to the attitude 
file. The first stage of the 
pipeline corrects the positions in the attitude file 
using {\tt attjumpcorr} and then recomputes the 
positions of every photon in the event list
using the corrected attitude file and the FTOOL 
{\tt coordinator}. The pipeline then refines the 
aspect of the event list by creating images 
every $10 {\rm ~s}$ and applying aspect correction 
software to each image. The aspect correction software 
locates the stars in each image and 
compares their positions to those in the USNO-B1 
catalogue, correcting for proper motions. 
For each image, it computes the RA and 
DEC offset between the stars 
in the image and those in the USNO-B1 catalogue. The 
offset is converted into pixels and then is 
applied to the position of each 
photon in the event list during the time interval of 
the image. To extract the photometry the 
pipeline runs {\tt uvotevtlc} on the 
aspect corrected event lists using $10 {\rm ~s}$ 
uniform binning, the $5\arcsec$ source regions and the background 
regions used to extract the image photometry.

The event lists, like the images, can suffer from 
a number of sources of contamination or degradation. These 
sources have been identified and 
flagged in a similar fashion as the UVOT images.

\subsection{Catalog Construction}
The UVOT GRB Catalog was constructed by combining 
information from various sources. Burst positions, 
trigger time, and time to the first UVOT observation 
were extracted from the image database. The magnitude 
of the first and peak detections were also determined 
for each filter. A value of ``-99" indicates that 
no UVOT data was available, while a value of ``99" 
indicates a $<3\sigma$ detection. 

Additional information in the 
catalog was gleaned from the literature. A reference 
to the best reported burst position is provided. 
Also included is a flag indicating which observatory 
discovered each burst. Galactic absorption and HI 
column density along the line of sight, $T_{90}$, 
redshift, GRB fluence 
in the $15-150 {\rm ~keV}$ band, radio flux, and a 
flag to indicate detections in ground-based $R-K$ 
bands are provided for each burst. Temporal slopes are derived for 
bursts with a sufficient number of significant detections, from 
which magnitudes are computed at $2000 {\rm ~s}$ (see Section 5). 

\subsection{Quality Control}
To verify the quality of the data, the following 
checks were performed: photometric and astrometric
stability of field stars, light curve
production and investigation of previously detected and 
non-detected afterglows, comparison of photometry to
previously published results, and visual examination of flagged
images. A description of each quality check is found
below. 

The stability of photometric and astrometric measurements made by the
pipeline were tested by applying the pipeline to stars, located
in the UVOT GRB fields, which have reliable astrometric and
photometric measurements from the Sloan Digital Sky Survey 
\citep[SDSS;][]{YDG00} database. From this database 108 test stars in 32
GRB fields were selected which were: within $2\arcmin$ of the 
GRB location, have magnitudes such that the stars are detectable in the UVOT
images, and are detected at the $3\sigma$-level. Since 
we can not select where GRBs are located, the
field stars were not selected to be standard stars. However, 
obvious variable stars were rejected from the sample.
Astrometric and photometric measurements were made of these
stars in every UVOT image that covers their locations. Statistical
characterization of the distributions of positions and count rates of the
stars was done in order to quantitatively describe both the internal
and absolute accuracy and precision of the pipeline measurements. 

\begin{figure}
\epsscale{0.9}
\plottwo{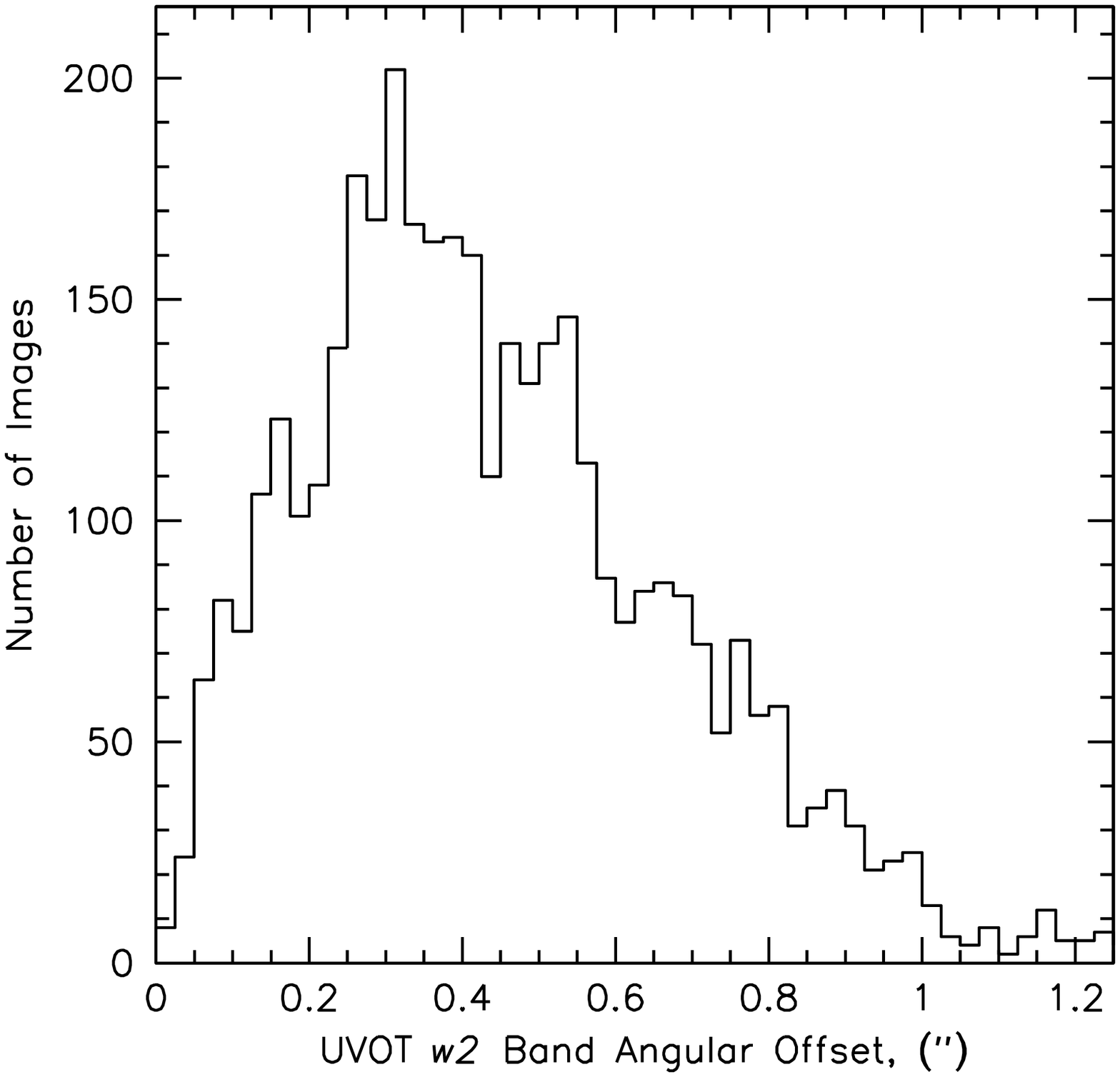}{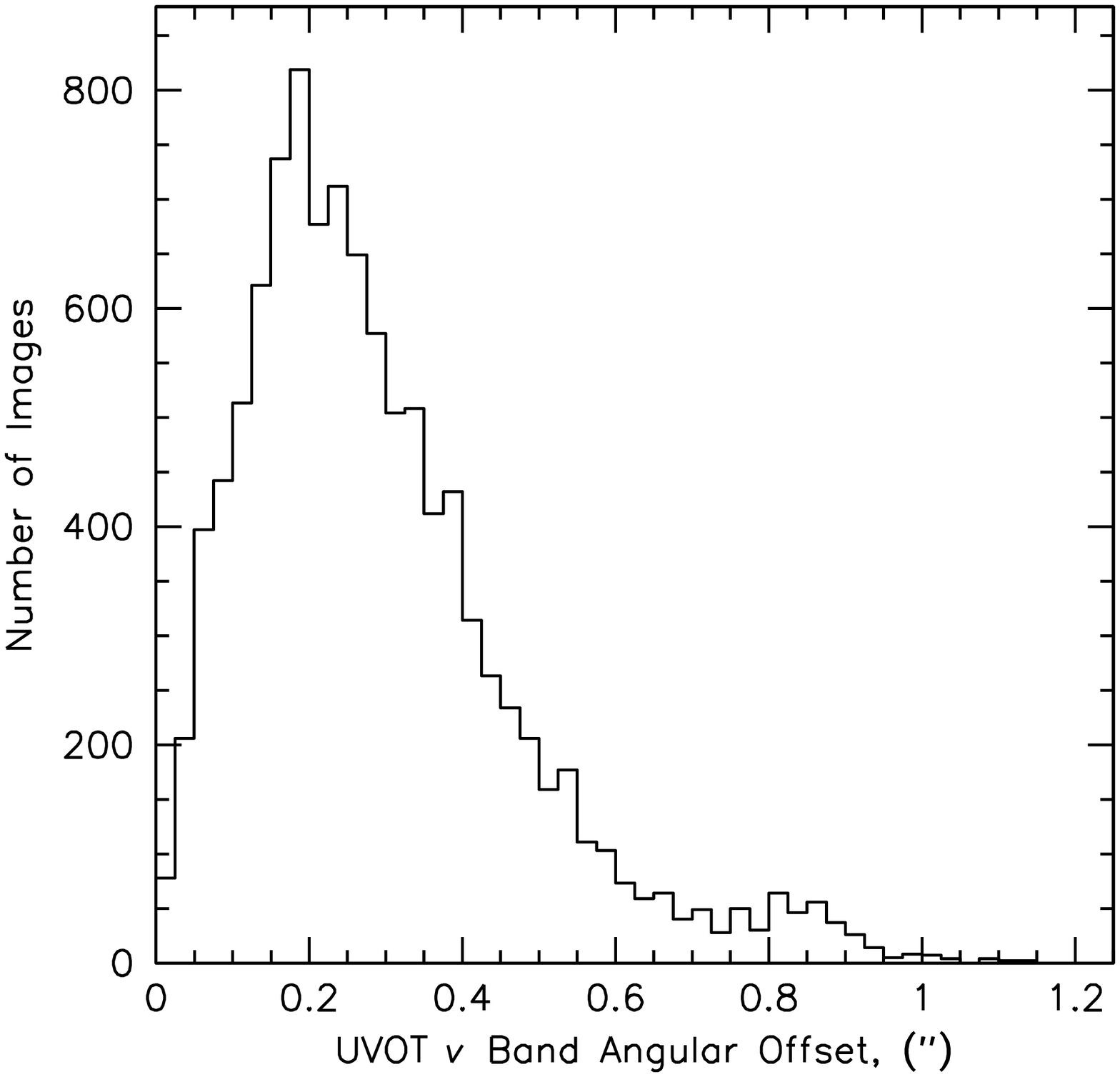}
\caption{Distribution of angular offsets with
respect to the USNO-B1 catalog for field stars observed
with the uvw2 ({\em left}) and $v$ ({\em right}) filters. 
The absolute accuracy is $0\farcs31$ and 
$0\farcs19$ for the uvw2 and $v$ filters, respectively.
The internal astrometric precision is $0\farcs27$ and $0\farcs21$
for the uvw2 and $v$ bands respectively.}
\label{fig-astrometric}   
\end{figure}

The mean position offsets in each band, relative to the USNO-B1 positions, defines
the absolute accuracy of the astrometry. The USNO-B1 positions are used
because the USNO-B1 catalog covers the entire sky and all UVOT
positions are determined from this catalog\footnote{It has been noted by
\citet{MDG03} that there is a systematic
offset as large as $0\farcs25$ between the SDSS and USNO-B1 positions
{\em after} correcting for proper motions.}. The peak of the angular 
offset is $0\farcs31$, $0\farcs41$, $0\farcs31$, $0\farcs19$, 
$0\farcs22$, $0\farcs19$, and $0\farcs14$ for the uvw2, uvm2, uvw1,
$u$, $b$, $v$, and $white$ filters, respectively (cf. 
Figure~\ref{fig-astrometric}). To determine the internal precision 
of the astrometry, we have calculated
the Rayleigh scale parameter for the distributions of angular offsets
from the mean stellar positions. A Rayleigh distribution holds if the
offsets in RA and DEC are independent and normally distributed with the
same standard deviation, which is then equivalent to the Rayleigh scale
parameter.  The internal astrometric precision (given by the Rayleigh
scale parameter) of each of the UVOT bands is $0\farcs27$, $0\farcs28$,
$0\farcs24$, $0\farcs22$, $0\farcs21$, $0\farcs21$, and $0\farcs17$
for the uvw2, uvm2, uvw1, $u$, $b$, $v$, and $white$ bands respectively.

\begin{figure}
\epsscale{0.9}
\plotone{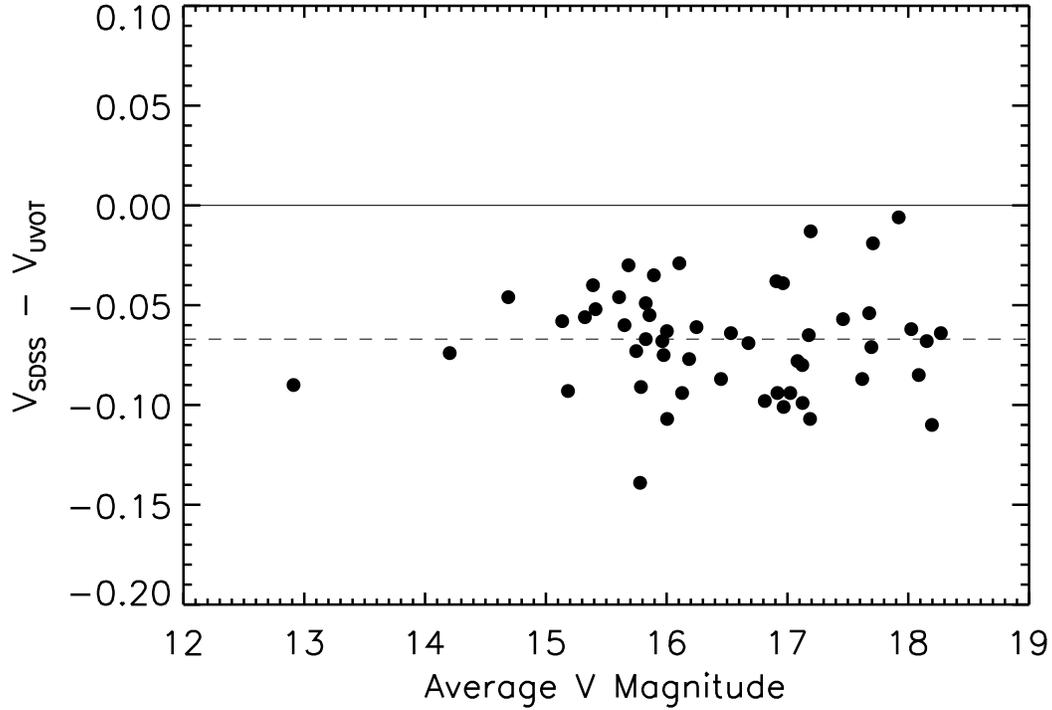}
\caption{The absolute photometric accuracy for the
$v$ filter. Comparing stars from the SDSS database with
UVOT observations and adjusting to the Johnson V
magnitude produces a median 
absolute photometric offset (dotted line) between the SDSS and UVOT
for the $v$ filter of 
-0.068 magnitudes ($3\sigma$ confidence limit).}
\label{fig-color}   
\end{figure}

\begin{figure}
\epsscale{0.9}
\plottwo{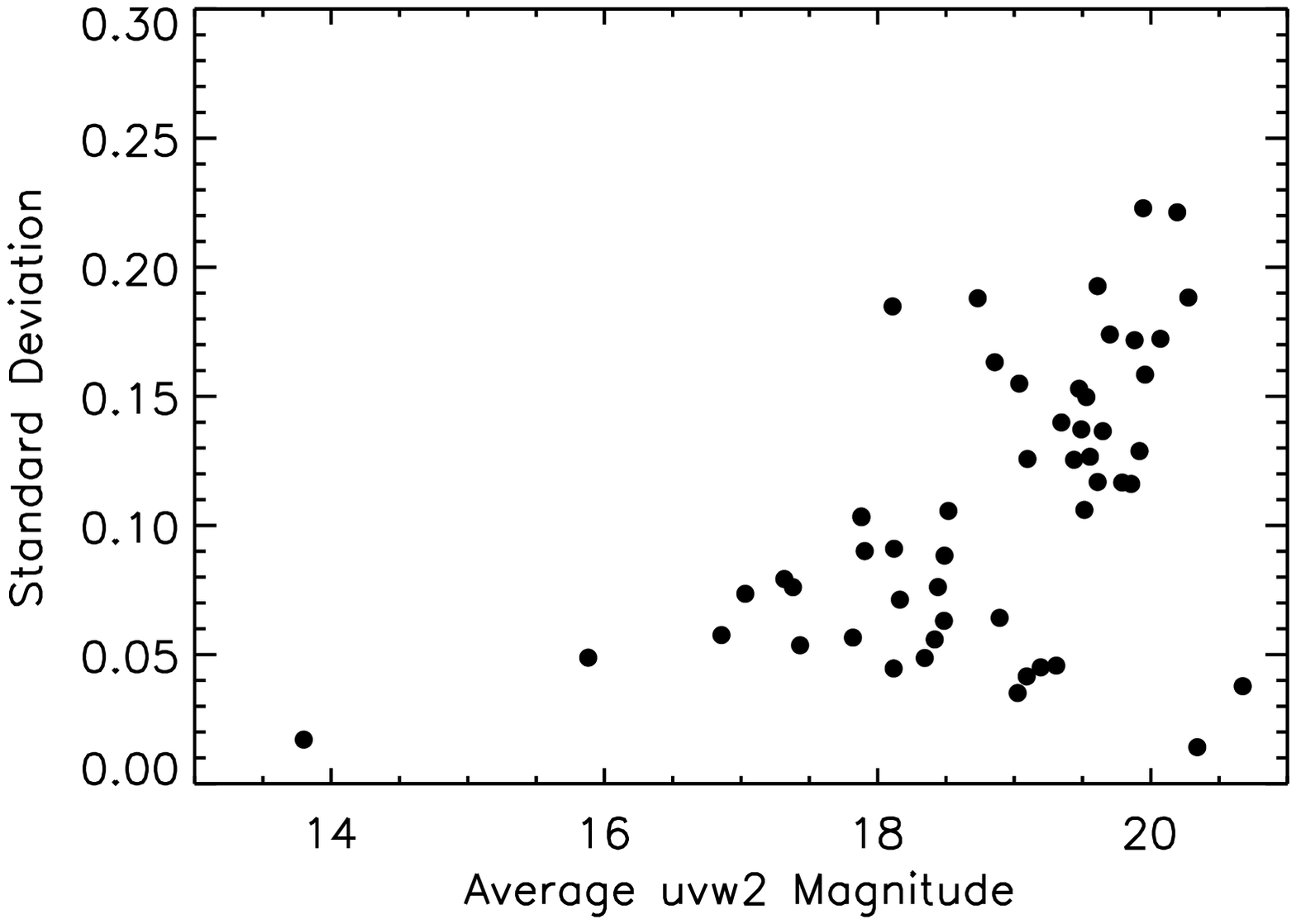}{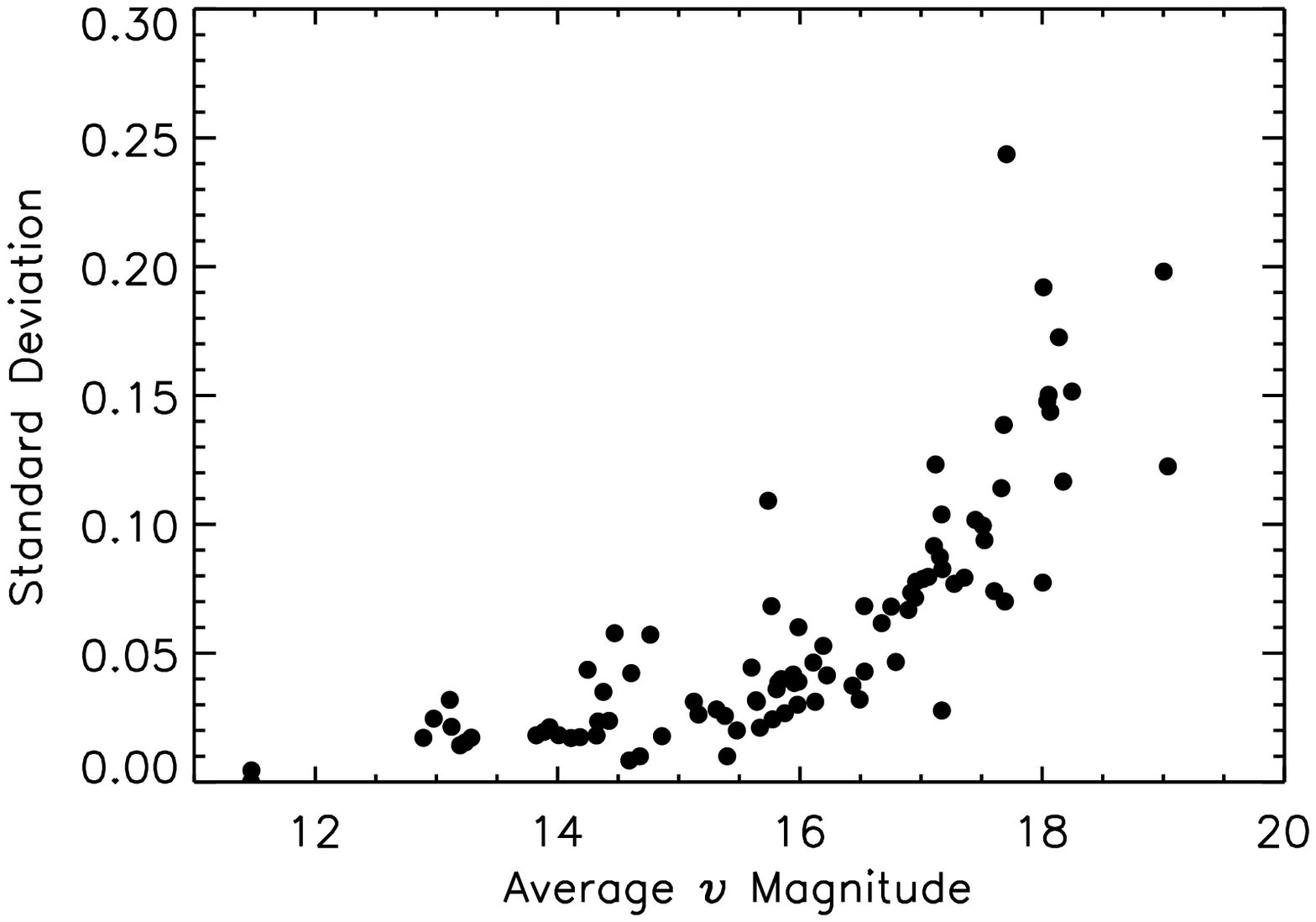}
\caption{Internal photometric precision for the uvw2 
({\em left}) and $v$ ({\em right}) filter data sets
for exposure times greater than $90 {\rm ~s}$.}
\label{fig-photometric}   
\end{figure}

The mean count rates of the stars, converted to magnitudes on the Johnson
system, is compared with the SDSS magnitudes transformed to the Johnson
system, to define the absolute accuracy of the photometry. The conversions
from SDSS and UVOT magnitudes to the Johnson system are based on work
by \citet{JS05} and \citet{PTS2007}, respectively. The average 
absolute photometric offset between the SDSS and UVOT 
(${\rm SDSS-UVOT}$) is $+0.076$ ($\sigma=0.052$), 
$+0.010$ ($\sigma=0.049$), and $-0.068$ ($\sigma=0.027$) magnitudes 
($3\sigma$ confidence limit) for the 
$u$, $b$, and $v$ filters, respectively (cf. Figure~\ref{fig-color}). The standard
deviation of the mean count rates defines the internal precision
of the photometry. The average standard deviation about the mean, 
over the nominal magnitude range of $13.8-20.7$ is 
0.11, 0.13, and 0.09 mag for the uvw2, uvm2, 
and uvw1 filters, respectively, and over the nominal magnitude
range of $11.5-19.4$ is 0.06, 0.05, 0.06, and 
0.05 mag for the $u$, $b$, $v$, and $white$ filters, respectively 
(cf. Figure~\ref{fig-photometric}).

We also compared the catalog entries to published values.  
In a small number of cases, no sources were detected 
in the database whereas the literature provides light curves for 
faint detections. This is because the photometry 
database is constructed with individual, not co-added, 
images. The UVOT burst database results were also compared
against the following published results in order to perform a 
consistency check for the following detected burst afterglows:
GRBs 060218 \citep{CS06}, 050525A 
\citep{BAJ06}, 060313 \citep{RPWA06}, 
061007 \citep{SP07}, 050319 \citep{MKO06}, 050318 \citep{SM05}, 
050603 \citep{GD06}, 
051117A \citep{GM07}, 050801 \citep{DM07}, 050730 
\citep{PM07}, and 050802 \citep{OS2007}. The comparison
resulted in all the catalog data being consistent with
the published data.

Images will still be in the UVOT catalog even if certain
quality checks are not passed. Some of the bad aspect 
correction flags were determined by keywords in the image 
files themselves, while others were flagged by the 
{\tt uvotphot} photometry
pipeline. Below is a description of the checks performed 
on all flagged images.

{\it Settling}: When the spacecraft reaches
the target location, it requires a brief
period of time to lock onto the target;
this period is known as settling. During this 
time the UVOT is observing in event mode,
typically in the $v$-filter. Because the UVOT 
detector voltage is changing during this period, the
count rate is not calibrated; therefore,
these images are flagged as settling
exposures and should be used with caution. 
The criteria for flagging an image as a
settling exposure is if the image is: the first in the
first observation sequence, taken in event mode,
taken in the $v$-filter, and the exposure time
is $<11 {\rm ~s}$.

Each GRB observation for which the first finding chart
exposure is present should include a settling exposure just
prior to the start of the observation. This condition was
checked in the database file, and cases where no settling
image was flagged were investigated; this ensures
that all settling images are flagged as such.
To ensure that non-settling images have not been
erroneously flagged, any settling mode images that 
were found to occur {\it after} the
initial finding chart exposure were also investigated. 

{\it Aspect Correction}: A small number of images which 
have a successful fine aspect correction are blurred due 
to uncorrected movement of the spacecraft. These images
have been flagged. 

{\it Charge Trailing or In Halo}: A subset of the database 
have been generated by filtering for anything flagged as 
``charge trailing" or ``in
halo." Individual fields from this list were visually inspected to
verify the condition. A random check in individual fields was
also made in order to check for the existence of these conditions
in cases where they were not flagged. 

{\it Edge Effect or Crowded Field}: Some images are not well centered 
on the burst and suffer from ``edge effects", while other 
bursts are embedded in crowded star fields. These images
were flagged.

{\it Near Bright Star}: Co-added images in $b$ and uvw1 from the initial
snapshot in each field were visually examined to search for
nearby bright stars. Nearby bright stars with diffraction spikes 
impinging on the source aperture were flagged.

{\it Cumulative Quality}: If any of the quality flags above are
set to true (``T") this flag is also set to true. These images
should be treated with caution when used in a
dataset. Approximately $10\%$ of all images are
flagged. The largest contribution to flagged images are
a result of in halo and charge trailing, totaling
$\sim6.2\%$ and $\sim1.3\%$ of the images, respectively.

\section{Database and Catalog Formats}
The {\em Swift} UVOT Image Mode Burst Database
(sample columns and rows are provided in Table~\ref{tab1}), the {\em Swift} 
UVOT Event Mode Burst Database (sample columns and rows are 
provided in Table~\ref{tab7}), and the {\em Swift} UVOT Burst 
Catalog (sample columns and rows provided in Table~\ref{tab2}) can be 
found at the {\em Swift} website. The databases and catalog are 
available in the following file formats: (1) a 
standard ASCII file with fixed column widths 
(size = $12.3 {\rm ~MB}$, $521 {\rm ~kB}$, \& $21 {\rm ~kB}$ for the image and
event databases, and the catalog, respectively) and (2) a binary FITS 
table (size = $13.9 {\rm ~MB}$, $560 {\rm ~kB}$, \& $28 {\rm ~kB}$ for the image and
event databases, and the catalog, respectively).

The Image Mode Database contains 
86 columns and 63,315 rows. Each 
column is described in Table~\ref{tab3}. Except 
for the object ID, time of burst trigger, filter, 
quality flags, name
of the FITS extension, trigger number,
and filename in columns 1, 6, 12, and 76-86, 
all entries are in floating point,
exponential, or integer format.

The Event Mode Database contains 
41 columns and 9402 rows. Each 
column is described in Table~\ref{tab8}. Except 
for the object ID, time of burst trigger, filter, quality flags, 
trigger number, and filename in columns 1, 6, 12, and 38-41, 
all entries are in floating point,
exponential, or integer format.

The Burst Catalog contains 
81 columns and 229 rows. Each 
column is decribed in Table~\ref{tab4}. Except 
for the object ID, position reference, time of burst trigger,
detection flags, and the notes in 
columns 1, 5, 9, and 78-81, 
respectively, all entries are in floating point,
exponential, or integer format.

Below are the notes on the image database columns found in 
Table~\ref{tab3}. Each note identifies the 
nomenclature and description of each column.\\

1. OBJECT: The object identification. The format is 
GRB$yymmddX$, where $yy$ is the last two digits of 
the year of the burst, $mm$ is the month, $dd$ 
is the day (in UTC), and $X$ is used to represent a 
second, third, fourth, etc., burst occuring on a 
given day by the letters `B' or `C'. 
Only the last seven characters are listed in the 
catalog (i.e. the ``GRB'' is dropped from each 
entry).

2. RA: The best J2000.0 right ascension, 
in decimal degrees, as found in 
the GCN
Circulars\footnote{http://gcn.gsfc.nasa.gov/gcn3\_archive.html},
\citet{GMR07}, and \citet{BNR07}. 
Thirteen GRBs have improved positions that 
were calculated from the centroid of the summed
images in the UVOT image database; these
have been identified in column 81 of 
the catalog.

3. DEC: The best J2000.0 declination, 
in decimal degrees, as described in RA above.

4. POS\_ERR: Positional uncertainty, in arcseconds,
as described in RA above.

5. TRIGTIME: The time of the burst trigger as measured in 
{\em Swift} mission elapsed time (MET). MET is 
measured in seconds and starts on 2001 January 1, 
00:00:00.000 (UTC). The MET of the launch of
{\em Swift}, 2004 November 20, 
18:35:45.865 (UTC), is 122668545.865.

6. TRIG\_UT: The time of the burst trigger as measured in 
Universal time (UTC) (e.g. 2005-017-12:52:36).

7. TIME: TSTART + TELAPSE/2 (see columns 8 \& 11).

8. TSTART: The MET start time of the exposure.

9. TSTOP: The MET stop time of the exposure.

10. EXPOSURE: The exposure time, in seconds, 
including the following corrections:
detector dead time, time lost when 
the on-board shift-and-add algorithm tosses
event data off the image, time lost when the
UVOT Digital Processing Unit stalls because
of high count rates, and time lost due to
exposures beginning with the UVOT blocked
filter.

11. TELAPSE: TSTOP - TSTART, in seconds.

12. FILTER: The UVOT filter used for the 
exposure (uvw2, uvm2, uvw1, $u$, $b$, $v$, 
and $white$).

13. BINNING: The binning factor ($1=1\times1 
{\rm ~binning}$ and $2=2\times2 {\rm ~binning}$).

14. APERTURE: The source aperture radius, in 
arcseconds.

15. SRC\_AREA: The area of the source region,
in square arcseconds, 
computed by multiplying the number of pixels 
found by XIMAGE within the source radius by 
the area of each pixel. This value can differ 
from the specified area $\pi r^2$ by up to $2\%$ 
because XIMAGE selects whole pixels within 
the source radius. This approach produces an 
area slightly larger or smaller than $\pi r^2$. 
Simulations reveal
that the $1\sigma$ difference between the 
exact and XIMAGE areas are $1.0\%$ and $1.5\%$
for a 10 and 6 pixel radius, respectively.
The error in photometry is much less 
than these area fluctuations because source 
counts are concentrated in the center of 
the aperture and the aperture correction 
uses the radius corresponding to the XIMAGE 
area. 

16. BKG\_AREA: The area, in square arcseconds, 
of the background region. It is calculated by
taking the number of pixels in the background 
annulus and multiplying by the area of each 
pixel. Masked regions are excluded therefore
only net pixels are included.
This differs slightly from the exact 
area $\pi(r_o-r_i)^2$, but we are only 
interested in the background surface 
brightness, so the difference is not significant.

17. PLATE\_SCALE: The plate scale, in arcseconds 
per pixel, of the image. The error in the mean
plate scale is $\pm 0\farcs0005 {\rm ~pixel^{-1}}$.

18. RAW\_TOT\_CNTS: Total number of counts 
measured within the source region.

19. RAW\_TOT\_CNTS\_ERR: The binomial 
error in RAW\_TOT\_CNTS. The binomial error is 
given by ${\rm (RAW\_TOT\_CNTS)}^{1/2}$ * 
${\rm ((NFRAME - RAW\_TOT\_CNTS)/NFRAME)}^{1/2}$. 
NFRAME = TELAPSE / FRAMETIME, where ${\rm FRAMETIME} = 
0.011032 {\rm ~s}$ for the full FoV. NFRAME
is the number of CCD frames (typically one
every $\sim11 {\rm ~ms}$). A discussion
of the measurement errors in the UVOT can be
found in \citet{KR08}.

20. RAW\_BKG\_CNTS: Total number of counts 
measured in the background annulus.

21. RAW\_BKG\_CNTS\_ERR: The binomial 
error in RAW\_BKG\_CNTS. The binomial error is 
given by ${\rm (RAW\_BKG\_CNTS)}^{1/2}$ * 
${\rm ((NFRAME - EFF\_BKG\_CNTS)/NFRAME)}^{1/2}$. 
EFF\_BKG\_CNTS = RAW\_BKG\_CNTS * 80 / 
BKG\_AREA. The effective counts in the 
background (EFF\_BKG\_CNTS) is calculated
because the background area is larger than
the coincidence region. The value 80 is the
area (in square arcseconds) of our circular
aperture with a radius of $5\arcsec$.

22. RAW\_STD\_CNTS: Total number of counts 
measured within the standard $5\arcsec$ aperture. 
This constant value is based on the size of the 
current calibration aperture.

23. RAW\_STD\_CNTS\_ERR: Binomial error associated 
with RAW\_STD\_CNTS.

24. RAW\_TOT\_RATE: The total measured count 
rate, in counts per second, in the source region.
Calculated using RAW\_TOT\_CNTS / EXPOSURE.

25. RAW\_TOT\_RATE\_ERR: RAW\_TOT\_CNTS\_ERR 
/ EXPOSURE.

26. RAW\_BKG\_RATE: The total measured count rate,
in counts per second per square arcsecond, in the 
background region. Calculated using RAW\_BKG\_CNTS 
/ EXPOSURE / BKG\_AREA.

27. RAW\_BKG\_RATE\_ERR: RAW\_BKG\_CNTS\_ERR 
/ EXPOSURE / BKG\_AREA.

28. GLOB\_BKG\_RATE: The global background rate,
in counts per second per square arcsecond. The 
global background of each image is modeled as 
a Gaussian distribution. An iterative ``Sigma 
Clipping" is performed to eliminate contributions 
from field stars above the $3\sigma$ level. The
global background is then reported as the 
arithmetic mean of the clipped distribution 
along with the number of samples in the clipped 
distribution. In images where the background is 
well sampled, the ratio of local to global 
background rates is a good indicator of sources 
embedded in the halo of a nearby bright star. 

29. GLOB\_BKG\_AREA: The area, in square arcseconds, 
of the global background region.  

30. RAW\_STD\_RATE: The total measured count rate, 
in counts per second, in the coincidence loss 
region. Calculated using RAW\_STD\_CNTS / 
EXPOSURE.

31. RAW\_STD\_RATE\_ERR: RAW\_STD\_CNTS\_ERR 
/ EXPOSURE.

32. COI\_STD\_FACTOR: The coincidence-loss 
correction factor for the coincidence-loss 
region. This is calculated as follows. First, 
the COI\_STD\_RATE (which is not recorded) is 
calculated using the theoretical 
coincidence loss formula and the polynomial correction to
RAW\_STD\_RATE = RAW\_STD\_CNTS / EXPOSURE 
\citep[see eq. 1-3 in][]{PTS2007}. The value
COI\_STD\_FACTOR is then the ratio 
COI\_STD\_RATE / RAW\_STD\_RATE.

33. COI\_STD\_FACTOR\_ERR: The uncertainty 
in the coincidence correction 
\citep[see eq. 4 in][]{PTS2007}. 

34. COI\_BKG\_FACTOR: The coincidence-loss 
correction factor for the background region.

35. COI\_BKG\_FACTOR\_ERR: The uncertainty 
in the coincidence correction of the background counts within
the source aperture.

36. COI\_SRC\_CNTS: The coincidence-loss corrected 
counts in the source region. Calculated using 
(RAW\_TOT\_CNTS - (RAW\_BKG\_CNTS * SRC\_AREA /
BKG\_AREA)) * COI\_STD\_FACTOR.

37. COI\_SRC\_CNTS\_ERR: The error associated 
with COI\_SRC\_CNTS. Calculated using 
(RAW\_TOT\_CNTS\_ERR - (RAW\_BKG\_CNTS\_ERR * SRC\_AREA /
BKG\_AREA)) * COI\_STD\_FACTOR\_ERR.

38. COI\_BKG\_CNTS: The coincidence-loss corrected 
counts in the background region. Calculated using 
RAW\_BKG\_CNTS * COI\_BKG\_FACTOR.

39. COI\_BKG\_CNTS\_ERR: The error associated 
with COI\_BKG\_CNTS. Calculated using 
RAW\_BKG\_CNTS\_ERR * COI\_BKG\_FACTOR\_ERR.

40. COI\_TOT\_RATE: The coincidence-loss corrected 
raw count rate, in counts per second, in the source
region. Calculated using RAW\_TOT\_RATE * 
COI\_STD\_FACTOR.

41. COI\_TOT\_RATE\_ERR: Error in the COI\_TOT\_RATE
= RAW\_TOT\_RATE\_ERR * COI\_STD\_FACTOR.

42. COI\_BKG\_RATE: The coincidence-loss corrected 
background surface count rate, in counts per second per square 
arcsecond. Calculated 
using\\
RAW\_BKG\_RATE * COI\_BKG\_FACTOR.

43. COI\_BKG\_RATE\_ERR: Error in coincidence 
corrected background surface brightness. Calculates
using RAW\_BKG\_RATE\_ERR * COI\_BKG\_FACTOR.

44. COI\_SRC\_RATE: Coincidence corrected net 
count rate, in counts per second. Calculated 
using COI\_TOT\_RATE - COI\_BKG\_RATE * SRC\_AREA.

45. COI\_SRC\_RATE\_ERR: Error in the coincidence 
corrected net count rate. The errors in the 
source rate and the background rate are added in 
quadrature:\\ 
${\rm (COI\_TOT\_RATE\_ERR^2 + 
(COI\_BKG\_RATE\_ERR * SRC\_AREA)^2)^{1/2}}$. 

46. AP\_FACTOR: Aperture correction for going 
from a $3\arcsec$ radius to a $5\arcsec$ radius 
aperture for the $v$ filter.  This is computed 
using the PSF stored in the CALDB by 
{\tt uvotapercorr}. This is always set to 
1.0 unless the {\tt CURVEOFGROWTH} method is used. 
The source radius is defined to be 
$({\rm SRC\_AREA}/\pi)^{1/2}$, so that one 
uses an effective source radius to the actual 
pixel area used by XIMAGE.

47. AP\_FACTOR\_ERR: The $1\sigma$ error in
AP\_FACTOR. AP\_FACTOR\_ERR = \\
AP\_COI\_SRC\_RATE\_ERR / COI\_SRC\_RATE\_ERR.

48. AP\_SRC\_RATE: Final aperture and 
coincidence loss corrected count rate used 
to derive the flux and magnitudes. 
Calculated using AP\_FACTOR * COI\_SRC\_RATE.

49. AP\_SRC\_RATE\_ERR: Error on the final 
count rate. Calculated using\\ 
$({\rm COI\_SRC\_RATE\_ERR^2 + (fwhmsig * 
COI\_SRC\_RATE)^2})^{1/2}$. The
``fwhmsig" parameter is the fractional RMS variation 
of the PSF which is set to $3\arcsec$. This
variation is propagated through the uncertainty calculation, and is
added in quadrature to the corrected measurement uncertainty.

50. MAG: The magnitude of the source in the 
UVOT system computed from \\ AP\_COI\_SRC\_RATE. 
The value is set to 99.00 for upper-limits.

51. MAG\_ERR: The one-sigma error in MAG.
Unless otherwise specified, all 
errors are the 1-sigma statistical errors based 
on Poisson statistics. The value is set to
9.90E+01 if MAG was an upper limit.

52. SNR: Signal-to-noise ratio calculated from
COI\_TOT\_RATE and COI\_BKG\_RATE. 

53. MAG\_BKG: The sky magnitude, in magnitudes per square
arcsecond, in the UVOT system computed from 
COI\_BKG\_RATE.

54. MAG\_BKG\_ERR: The one-sigma error in MAG\_BKG.

55. MAG\_LIM: The ``N"-sigma limiting magnitude 
in the UVOT system computed from the RAW quantities.

56. MAG\_LIM\_SIG: ``N" for MAG\_LIM, where ``N" 
is a chosen parameter. The database uses a value of
3 for N.

57. MAG\_COI\_LIM: The magnitude at which the count 
rate is one count per CCD frame.

58. FLUX\_AA: The flux density in 
${\rm ~erg\,s^{-1}\,cm^{-2}\,\AA^{-1}}$.

59. FLUX\_AA\_ERR: The one-sigma error in 
FLUX\_AA.

60. FLUX\_AA\_BKG: The flux density of the sky in 
${\rm ~erg\,s^{-1}\,cm^{-2}\,\AA^{-1}}$ per
square arcsecond.

61. FLUX\_AA\_BKG\_ERR: The one-sigma error in 
FLUX\_AA\_BKG.

62. FLUX\_AA\_LIM: The approximate flux density
limit based on an average GRB spectrum, 
in ${\rm ~erg\,s^{-1}\,cm^{-2}\,\AA^{-1}}$.

63. FLUX\_AA\_COI\_LIM: The flux density at which 
the count rate is one count per frame time, in
${\rm ~erg\,s^{-1}\,cm^{-2}\,\AA^{-1}}$.

64. FLUX\_HZ: The flux density in 
${\rm ~mJy}$.

65. FLUX\_HZ\_ERR: The one-sigma error in FLUX\_HZ.

66. FLUX\_HZ\_BKG: The flux density of the sky in 
${\rm ~mJy}$ per square
arcsecond.

67. FLUX\_HZ\_BKG\_ERR: The one-sigma error in 
FLUX\_HZ\_BKG.

68. FLUX\_HZ\_LIM: The ``N"-sigma limiting 
flux density in 
${\rm ~mJy}$, 
corresponding to MAG\_LIM.

69. FLUX\_HZ\_COI\_LIM: The flux density at which 
the count rate is one count per frame time, in
${\rm ~mJy}$.

70. NEAREST\_RA: The J2000.0 right ascension, 
in decimal degrees, of the closest non-GRB 
source within $15\arcsec$ to the burst position, as determined
by {\tt uvotdetect}.

71. NEAREST\_DEC: The J2000.0 declination, 
in decimal degrees, of the closest non-GRB 
source within $15\arcsec$ to the burst position, as determined
by {\tt uvotdetect}.

72. NEAREST\_MAG: The magnitude, in the given
UVOT band, of the object at
CLS\_SRC\_RA and CLS\_SRC\_DEC, as determined
by {\tt uvotdetect}. If there is no source
found within $15\arcsec$ radius of the 
burst position then this value is set to 99.00.

73. CENTR\_RA: The source's centroided right ascension, 
in decimal degrees, as determined by a 2D
gaussian fit of the UVOT data. If the fit failed
then the value is set to 999.000000.

74. CENTR\_DEC: The source's centroided declination, 
in decimal degrees, as determined by a 2D
gaussian fit of the UVOT data. If the fit failed
then the value is set to 999.000000.

75. CENTR\_ERR: The larger of the fit errors
between CENTR\_RA and CENTR\_DEC divided by the
square-root of the number of counts (N). Given as a
$1\sigma$ error. If the fit failed, then a 
value of $-1.0$ is assigned. N = MAX(1 , 
RAW\_TOT\_CNTS - [RAW\_BKG\_RATE * EXPOSURE *
SRC\_AREA]).

76. SETTLE\_FLAG: Settling images sometimes 
have a poor aspect solution which creates 
doublets out of field stars. Settling images 
also suffer from detector gain issues because 
the High Voltages may still be ramping up part 
way into the exposure. Such images have an 
undefined photometric calibration and should 
be used cautiously. Images fitting all of the following 
are flagged as settling exposures, i.e. this 
flag is true (T): first exposure 
of Segment 0, image taken in Event mode, and
${\rm EXPOSURE} < 11 {\rm ~s}$.

77. ASPECT\_FLAG: The {\em Swift} spacecraft pointing 
accuracy is $\approx5\arcsec$. The astrometric error is 
improved to about $0\farcs3$ by comparing source 
positions to the USNO-B catalog. For a small number 
of images the automated procedure did not produce
an aspect solution. Such images are flagged as true (T).

78. TRAIL\_FLAG: A number of $v$ and 
$white$ images suffer from the effects of charge 
trailing. This happens when bright sources align 
with the source aperture in the CCD readout 
direction. Visible bright streaks along CCD 
columns (RAWY) sometimes complicate photometric 
measurements. These images are flagged
true (T). 

79. CROWDED\_FLAG: If the field appears crowded
upon visual inspection the image is flagged 
true (T). 

80. SPIKE\_FLAG: If a diffraction spike impinges
upon the source region then the image is flagged 
true (T). 

81. EDGE\_FLAG: If the source is sufficiently close to
the edge of the image such that the exposure
across the background region is variable then
the image is flagged true (T). 

82. HALO\_FLAG: A few bursts lie within the 
halo of bright stars, which can produce erroneous 
photometric measurements. Such situations are
flagged by comparing the local background to 
the global background. These images are flagged
true (T). 

83. QUALITY\_FLAG: Cumulative quality flag. 
This flag is set to true (T) when any of the 
following
quality flags are true (T): SETTLE\_FLAG, 
ASPECT\_FLAG, TRAIL\_FLAG, SPIKE\_FLAG, 
EDGE\_FLAG, or HALO\_FLAG.

84. TRIG\_NUM: The {\em Swift} triggering
number for the burst. 

85. EXTNAME: The name of the FITS extension 
that contains this observation. The name of the extension
has the following format: \{filter\}\{exposure ID\}\{I/E\}
where \{filter\} = wh, vv, bb, uu, w1, m2, or w2 for
the $white$, $v$, $b$, $u$, uvw1, uvm2, and uvw2, 
respectively, and \{I/E\} represents image or
event mode (e.g. vv133535746I).

86. IMAGE\_NAME: Name of the image (e.g. \\
00020004001/uvot/image/st00020004001ubb\_sk.img.gz[1]).
\\

Below are the notes on the event database columns found in 
Table~\ref{tab8}. Each note identifies the 
nomenclature and description of each column.\\

1-17. OBJECT, RA, DEC, POS\_ERR, TRIGTIME, 
TRIG\_UT, TIME, TSTART, \\
TSTOP, EXPOSURE, TELAPSE, FILTER,
BINNING, APERTURE, SRC\_AREA, \\BKG\_AREA, \&
PLATE\_SCALE: 
Same as columns 1-17 in Table~\ref{tab3},
respectively.

18-21. COI\_STD\_FACTOR, COI\_STD\_FACTOR\_ERR,
COI\_BKG\_FACTOR, \& \\
COI\_BKG\_FACTOR\_ERR: 
Same as columns 32-35 in Table~\ref{tab3},
respectively.

22-28. COI\_TOT\_RATE, COI\_TOT\_RATE\_ERR, 
COI\_BKG\_RATE, \\
COI\_BKG\_RATE\_ERR, 
COI\_SRC\_RATE, COI\_SRC\_RATE\_ERR\footnote{The 
values for COI\_SRC\_RATE\_ERR \& AP\_SRC\_RATE\_ERR
are not calculated correctly by the pipeline 
software and are therefore too large. These values will be corrected in
future versions of the database.}, \&
AP\_FACTOR: 
Same as columns 40-46 in Table~\ref{tab3},
respectively.

29-32. AP\_SRC\_RATE, 
AP\_SRC\_RATE\_ERR, MAG, \& MAG\_ERR: 
Same as columns 48-51 in Table~\ref{tab3},
respectively.

33-34. FLUX\_AA \& FLUX\_AA\_ERR: 
Same as column 58-59 in Table~\ref{tab3},
respectively.

35-39. CENTR\_RA, CENTR\_DEC, CENTR\_ERR, SETTLE\_FLAG, 
ASPECT\_FLAG: 
Same as columns 73-77 in Table~\ref{tab3},
respectively. 

40-41. TRIG\_NUM \& FILENAME: Same as column 85-86 
in Table~\ref{tab3}.
\\

Notes on the {\em Swift} UVOT burst catalog 
columns found in Table~\ref{tab4}:\\

1-4. OBJECT, RA, DEC, \& PNT\_ERR: 
Same as columns 1-4 in Table~\ref{tab3},
respectively.

5. POS\_REF: The position reference for columns
2-4. References are from the the GCN Circulars,
\citet{GMR07}, and \citet{BNR07}. 

6. DISC\_BY: The ``discovery flag" indicating which 
spacecraft discovered the GRB. The flag is an 
integer from 0-2 representing: 0 = {\em Swift}, 1 = 
HETE2, 2 = INTEGRAL, and 3 = IPN.

7. T90: $T_{90}$, in seconds, as defined by \citet{ST2007}
for the {\em Swift} discovered bursts in the $15-350
{\rm ~keV}$ band. For GRBs 050408 \citep{ST05}, 
051021 \citep{OJ05}, 051028 \citep{HK05b}, 
051211A \citep{KN05}, and 060121 \citep{AM05}
discovered by HETE2, $T_{90}$ in the $30-400 
{\rm ~keV}$ band is provided as in the GCN Circulars 
($T_{90}$ for GRB 060121 is provided in the 
$80-400 {\rm ~keV}$ band). 
$T_{90}$ for the
INTEGRAL and IPN discovered bursts, as well as the remaining
HETE2 bursts, were not available in the GCN Circulars. 
In all cases where
$T_{90}$ was not available this value is set to $-99.0$. 

8-9. TRIGTIME \& TRIG\_UT: Same as columns 5 and 6 from 
Table~\ref{tab3}, respectively.

10. FRST\_TSTART: The start time of the first UVOT 
observation measured in seconds from the burst trigger.

11. FRST\_W2: The first reported magnitude in the
uvw2-filter. If no detections are reported then this
value is set to 99.00. If not observed in the
uvw2-filter then this value is set to $-99.00$.

12. FRST\_W2\_T: The time since burst, in seconds, 
to the middle of the exposure of FRST\_W2. If
FRST\_W2 = $\pm 99.00$ then this value is set to
-1.0.

13. FRST\_M2: Same as FRST\_W2 except for the 
uvm2-filter.

14. FRST\_M2\_T: The same as FRST\_W2\_T except
for the uvm2-filter.

15. FRST\_W1: Same as FRST\_W2 except for the 
uvw1-filter.

16. FRST\_W1\_T: The same as FRST\_W2\_T except
for the uvw1-filter.

17. FRST\_UU: Same as FRST\_W2 except for the 
$u$-filter.

18. FRST\_UU\_T: The same as FRST\_W2\_T except
for the $u$-filter.

19. FRST\_BB: Same as FRST\_W2 except for the 
$b$-filter.

20. FRST\_BB\_T: The same as FRST\_W2\_T except
for the $b$-filter.

21. FRST\_VV: Same as FRST\_W2 except for the 
$v$-filter.

22. FRST\_VV\_T: The same as FRST\_W2\_T except
for the $v$-filter.

23. FRST\_WH: Same as FRST\_W2 except for the 
$white$-filter.

24. FRST\_WH\_T: The same as FRST\_W2\_T except
for the $white$-filter.

25. PEAK\_W2: The peak reported magnitude in the
uvw2-filter. If no detections are reported then this
value is set to 99.00. If not observed in the
uvw2-filter then this value is set to $-99.00$.

26. PEAK\_W2\_T: The time since burst, in seconds, 
to the middle of the exposure of PEAK\_W2. If
PEAK\_W2 = $\pm 99.00$ then this value is set to
-1.0.

27. PEAK\_M2: Same as PEAK\_W2 except for the 
uvm2-filter.

28. PEAK\_M2\_T: The same as PEAK\_W2\_T except
for the uvm2-filter.

29. PEAK\_W1: Same as PEAK\_W2 except for the 
uvw1-filter.

30. PEAK\_W1\_T: The same as PEAK\_W2\_T except
for the uvw1-filter.

31. PEAK\_UU: Same as PEAK\_W2 except for the 
$u$-filter.

32. PEAK\_UU\_T: The same as PEAK\_W2\_T except
for the $u$-filter.

33. PEAK\_BB: Same as PEAK\_W2 except for the 
$b$-filter.

34. PEAK\_BB\_T: The same as PEAK\_W2\_T except
for the $b$-filter.

35. PEAK\_VV: Same as PEAK\_W2 except for the 
$v$-filter.

36. PEAK\_VV\_T: The same as PEAK\_W2\_T except
for the $v$-filter.

37. PEAK\_WH: Same as PEAK\_W2 except for the 
$white$-filter.

38. PEAK\_WH\_T: The same as PEAK\_W2\_T except
for the $white$-filter.

39. ALPHA\_W2: In the case of two or more afterglow 
detections in the uvw2-filter for a given burst, 
all occuring between $\sim300-100,000 {\rm ~s}$, 
the temporal slope ($\alpha_{uvw2}$) for that filter 
is provided. If two or more detections are not found 
for any given segment, the value is set to $-99.99$.
The value is calculated using,
\begin{equation}
f_{\lambda(uvw2)} = At^{-\alpha_{uvw2}},
\end{equation}
where $f_{\lambda(uvw2)}$ is the flux density, $A$ is 
the amplitude, and $t$ is the time since burst.

40. ALPHA\_W2\_ERR: The one-sigma error in ALPHA\_W2. If 
ALPHA\_W2 = $-99.99$ then ALPHA\_W2\_ERR = $-99.99$.

41. ALPHA\_W2\_AMP: The amplitude of ALPHA\_W2. If 
ALPHA\_W2 = $-99.99$ then ALPHA\_AMP = $-99.99$.

42. MAG\_ALPHA\_W2: The computed uvw2-filter 
magnitude at $2000 {\rm ~s}$ derived from using
ALPHA\_W2. If ALPHA\_W2 = $-99.99$ then 
MAG\_ALPHA\_W2 = $-99.99$.

43. ALPHA\_M2: Same as ALPHA\_W2 except for the 
uvm2-filter.

44. ALPHA\_M2\_ERR: Same as ALPHA\_W2\_ERR
except for the uvm2-filter.

45. ALPHA\_M2\_AMP: Same as ALPHA\_W2\_AMP
except for the uvm2-filter.

46. MAG\_ALPHA\_M2: Same as MAG\_ALPHA\_W2 except for the 
uvm2-filter.

47. ALPHA\_W1: Same as ALPHA\_W2 except for the 
uvw1-filter.

48. ALPHA\_W1\_ERR: Same as ALPHA\_W2\_ERR
except for the uvw1-filter.

49. ALPHA\_W1\_AMP: Same as ALPHA\_W2\_AMP
except for the uvw1-filter.

50. MAG\_ALPHA\_W1: Same as MAG\_ALPHA\_W2 except for the 
uvw1-filter.

51. ALPHA\_UU: Same as ALPHA\_W2 except for the 
$u$-filter.

52. ALPHA\_UU\_ERR: Same as ALPHA\_W2\_ERR
except for the $u$-filter.

53. ALPHA\_UU\_AMP: Same as ALPHA\_W2\_AMP
except for the $u$-filter.

54. MAG\_ALPHA\_UU: Same as MAG\_ALPHA\_W2 except for the 
$u$-filter.

55. ALPHA\_BB: Same as ALPHA\_W2 except for the 
$b$-filter.

56. ALPHA\_BB\_ERR: Same as ALPHA\_W2\_ERR
except for the $b$-filter.

57. ALPHA\_BB\_AMP: Same as ALPHA\_W2\_AMP
except for the $b$-filter.

58. MAG\_ALPHA\_BB: Same as MAG\_ALPHA\_W2 except for the 
$b$-filter.

59. ALPHA\_VV: Same as ALPHA\_W2 except for the 
$v$-filter.

60. ALPHA\_VV\_ERR: Same as ALPHA\_W2\_ERR
except for the $v$-filter.

61. ALPHA\_VV\_AMP: Same as ALPHA\_W2\_AMP
except for the $v$-filter.

62. MAG\_ALPHA\_VV: Same as MAG\_ALPHA\_W2 except for the 
$v$-filter.

63. ALPHA\_WH: Same as ALPHA\_W2 except for the 
$white$-filter.

64. ALPHA\_WH\_ERR: Same as ALPHA\_W2\_ERR
except for the $white$-filter.

65. ALPHA\_WH\_AMP: Same as ALPHA\_W2\_AMP
except for the $white$-filter.

66. MAG\_ALPHA\_WH: Same as MAG\_ALPHA\_W2 except for the 
$white$-filter.

67. BBMINUSVV: MAG\_ALPHA\_BB - MAG\_ALPHA\_VV. If 
MAG\_ALPHA\_BB or\\ MAG\_ALPHA\_VV is $-99.99$ 
the value for BBMINUSVV is also set to $-99.99$.

68. W1MINUSVV: MAG\_ALPHA\_W1 - MAG\_ALPHA\_VV. If 
MAG\_ALPHA\_W1 or\\ MAG\_ALPHA\_VV is $-99.99$ 
the value for W1MINUSVV is also set to $-99.99$.

69. RED: $E(B-V)$ as found in 
\citet{SFD98}. Galactic extinction can be corrected using 
the procedure described in 
\citet{CCM89}. The extinction in the UVOT bands
can be expressed as,
\begin{equation}
A_{\lambda} = E(B-V)[aR_v + b],
\end{equation}
where $R_v=3.1$ and $\lambda$ is the UVOT filter (uvw2, 
uvm2, uvw1, $u$, $b$, and $v$). The values for 
$a$ in each filter are $-0.0581$, 0.0773, 0.4346,
0.9226, 0.9994, and 1.0015, respectively. The
values for $b$ are 8.4402, 9.1784, 5.3286, 
2.1019, 1.0171, and 0.0126, respectively. All
values for $a$ and $b$ were determined as 
described in \citet{CCM89}. No
correction factor is provided for the $white$
filter as the large width of the FWHM makes
any extinction correction highly dependent on 
the spectral energy distribution of the source. 

70. NH: The logarithm of the absorption column 
density ($N_H$) along the line of sight as 
defined in \citet{KPMW05}.

71. ZZ: The redshift of the burst. If no redshift 
was found the value is set to 99.9999.

72. ZZ\_METH: The flag indicating how the redshift was 
determined. The flag 
is an integer from 0-4 representing: 0 = 
no redshift determined, 1 = absorption, 2 = emission, 
or 3 = Lyman break.

73. ZZ\_GCN: The GCN Circular number where the information
for ZZ and ZZ\_METH was reported. If no redshift was
reported this value is left blank.

74. FLUENCE\_BAT: The prompt BAT fluence of the burst,
in $10^{-8} {\rm ~erg ~cm^{-2}}$ ($15-150 {\rm ~keV}$ band), as reported
in \citet{ST2007}. If no fluence
was reported, or for HETE2, INTEGRAL, or IPN discovered
bursts, the value is set to $-99.0$.

75. FLUX\_RAD: The radio flux of the burst, in
${\rm mJy}$, as reported in the GCN Circular. If an
upper limit was reported the value is set to 99.000. 
If no radio observation was reported than the value
is set to $-99.000$.

76. FLUX\_RAD\_GCN: The GCN Circular number where the FLUX\_RAD
was reported. If no flux was
reported this value is left blank.

77. RADIO\_FREQ: The observed frequency of
the detection in FLUX\_RAD expressed in GHz.

78. DET\_IR: Flag indicating whether a detection in
the $R-K$ bands was reported in the GCN Circulars. 
``F" = No, ``T" = Yes.

79. DET\_UVOT: Flag indicating whether a detection in
any of the UVOT bands was found. 
``F" = No, ``T" = Yes.

80. DET\_RADIO: Flag indicating whether a detection in
the radio was reported in the GCN Circulars. 
``F" = No, ``T" = Yes.

81. NOTES: Notes on individual objects.

\section{Catalog Summary}
\begin{figure}
\epsscale{0.9}
\plotone{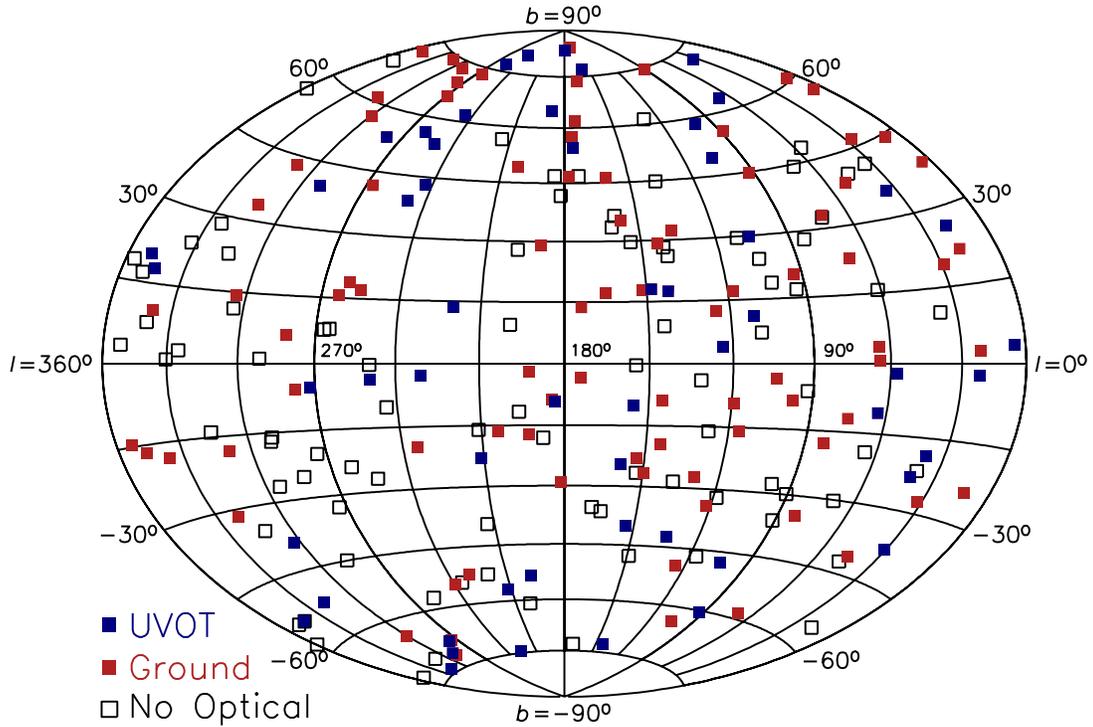}
\caption{Coordinates on a Galactic Aitoff projection of 
229 UVOT observed GRBs in the first $\sim2.5 {\rm ~years}$. Open 
squares are undetected afterglows by UVOT or ground-based 
telescopes, blue squares are afterglows detected 
by UVOT, and red squares are afterglows detected by 
ground based telescopes. [{\em See the electronic edition of
the Journal for a color version of this figure.}]}
\label{fig-aitoff}   
\end{figure}

From the catalog we can construct some of the general 
characteristics of the burst sample. 
Figure~\ref{fig-aitoff} displays the celestial coordinates of all UVOT 
observed GRBs, and whether or not they were detected. As is
expected, the distribution of all bursts is random on the sky 
\citep[cf.][]{MCA92}. From the sample of
the 229 observed GRBs, $\sim26\%$ are detected by the
UVOT at the $3\sigma$-level (in an individual exposure), $\sim60\%$ 
are detected by ground-based 
observations (although these are typically redder detections
than reported by the UVOT) as reported in the GCNs, and
$\sim40\%$ have no reported detections.

\begin{figure}
\epsscale{0.9}
\plotone{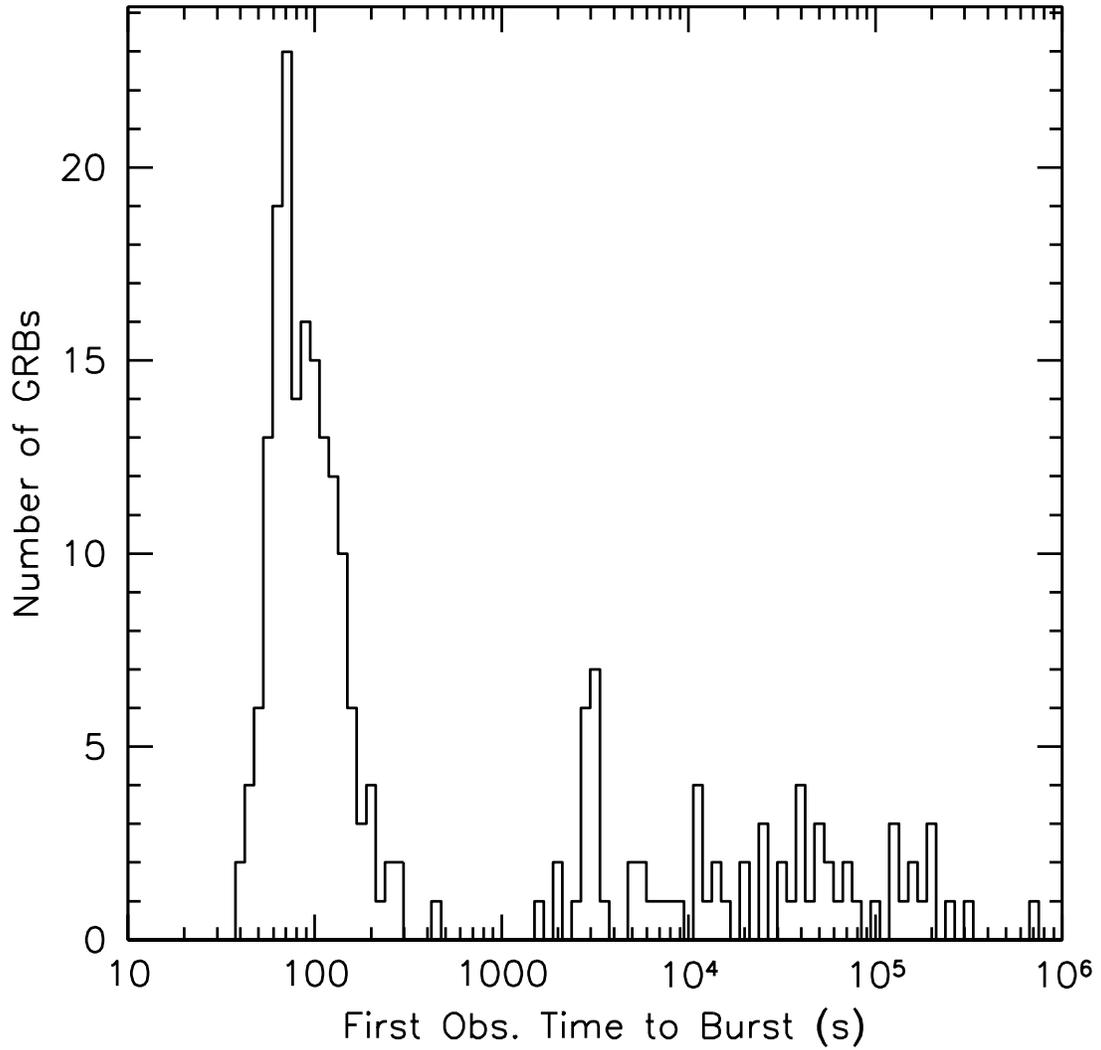}
\caption{The distribution of times for the first observation
of a GRB, including settling exposures. The median time to
burst observations is $110 {\rm ~s}$ for all observations. 
The median time to burst observations is $86 {\rm ~s}$ if
only the main peak is considered.}
\label{fig-firstobs}   
\end{figure}

The distribution of times to the start of the first observation 
(i.e. the settling exposure) by
UVOT is shown in Figure~\ref{fig-firstobs}. The median 
time to burst observation is $110 {\rm ~s}$ for all
observations. If only the main peak of the distribution
is considered, i.e. the $\sim70\%$ of the sample with 
$\Delta t < 300 {\rm ~s}$, the 
median time to burst observation is $86 {\rm ~s}$. 
In most cases the UVOT is observing the
very early stages of the afterglow. In some cases, UVOT
is observing during the end of the prompt emission
phase. The 
delay in observing the remaining $\sim30\%$ of bursts
is typically due to Earth occultations or to the 
inherent lag time for non-{\em Swift} burst alerts. 

\begin{figure}
\epsscale{0.9}
\plottwo{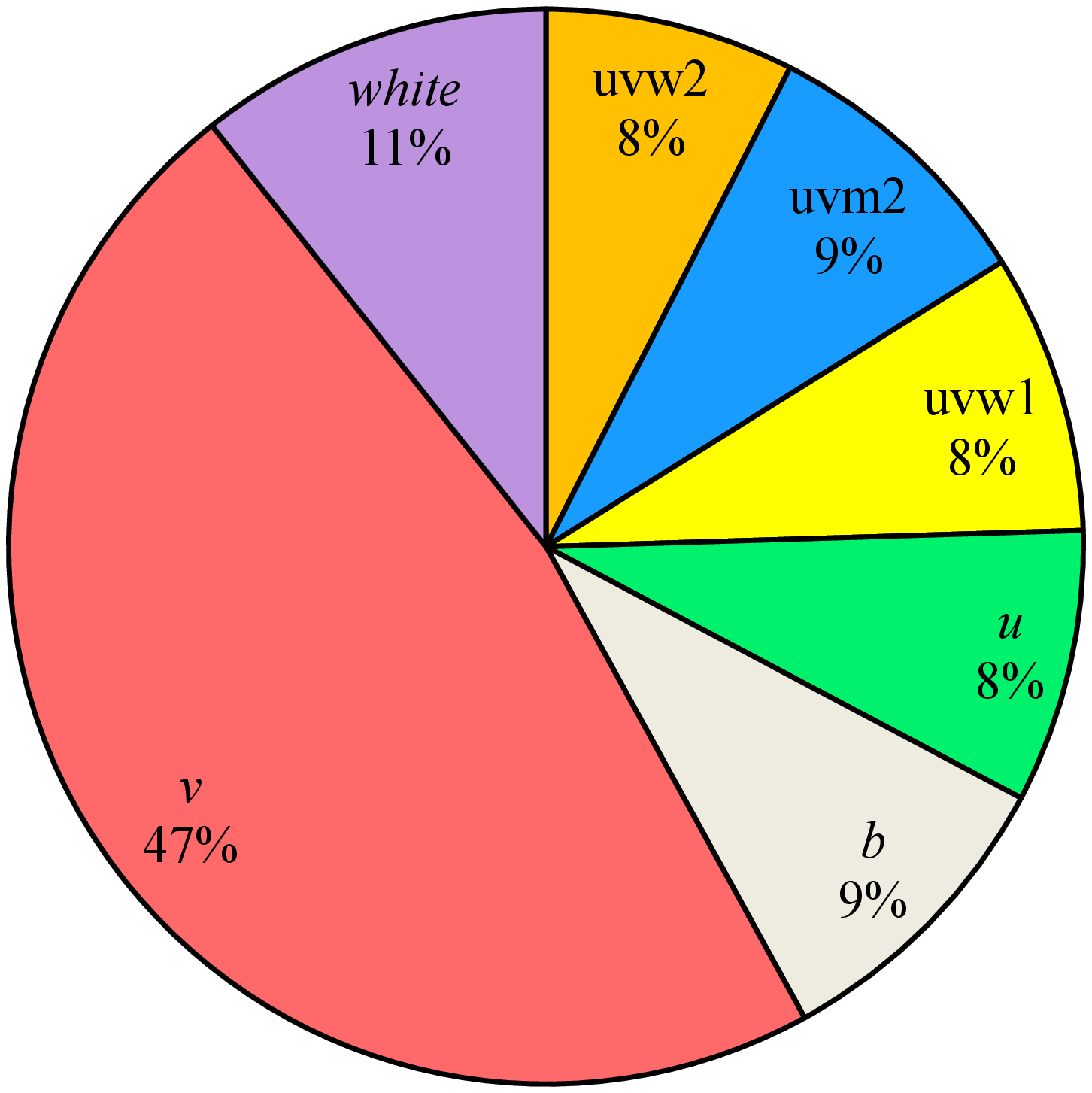}{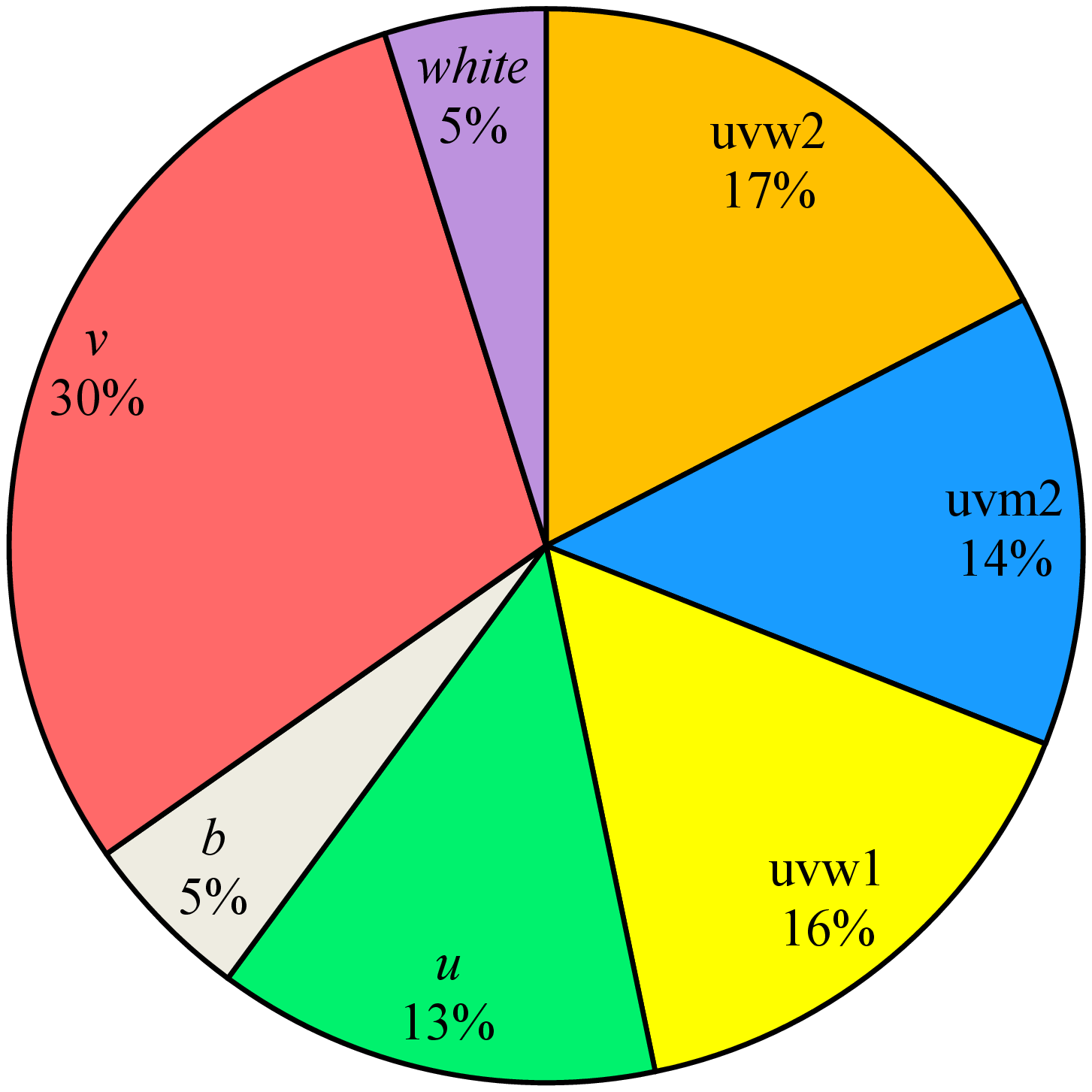}
\caption{The distribution of time spent in each
filter in observing
all bursts in the first $2000 {\rm ~s}$ ({\em left})
following the burst detection and for the cumulative time 
({\em right}) for all bursts. 
[{\em See the electronic edition of
the Journal for a color version of this figure.}]}
\label{fig-distribution}   
\end{figure}

Figure~\ref{fig-distribution} illustrates the distribution of
exposure times in each filter in the first $2000 
{\rm ~s}$ following the GRB detection, as well as the
total exposure in each filter for all bursts. 
Because the finding charts are typically taken in the
$white$ and $v$ filters, and since the finding charts
dominate the observing time during the first $\sim2000
{\rm ~s}$, the early light curves predominantly have
$white$ and $v$ data points.

\begin{figure}
\epsscale{0.9}
\plotone{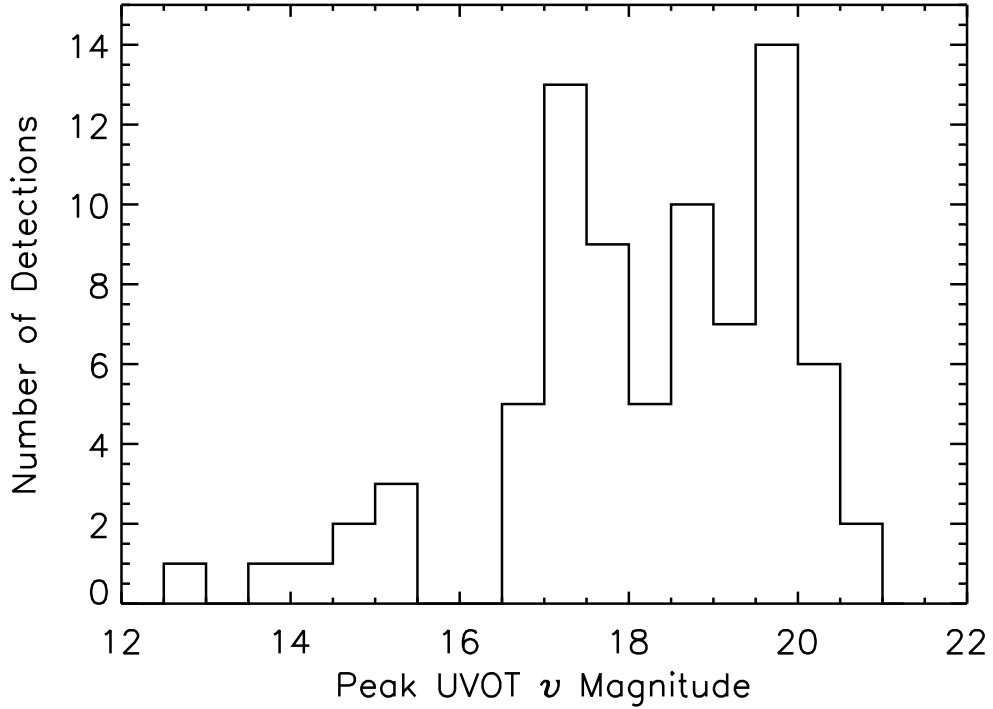}
\caption{Distribution of the brightest UVOT $v$-filter 
magnitudes for each detected burst for a total of 
79 points in the histogram. For time-to-observation 
of bursts $<500 {\rm ~s}$ and for galactic reddening
$<0.5$, the UVOT pipeline detects 
in a {\em single} exposure (no coadding of data) an afterglow 
in $\sim27\%$ of the cases. For time-to-observation of 
bursts $\geq500 {\rm ~s}$ and for Galactic reddening
$<0.5$, the UVOT pipeline detects 
in a {\em single} exposure an afterglow 
$\sim22\%$ of the time.}
\label{fig-peakmag}   
\end{figure}

The distribution of 
the brightest UVOT $v$-filter magnitudes for each 
detected burst, which is almost always the first or second 
exposure, is shown in Figure~\ref{fig-peakmag}. 
For time-to-observation 
of bursts $<500 {\rm ~s}$ and for Galactic reddening
$<0.5$ the UVOT pipeline detects 
in a {\em single} exposure (i.e. no coadding of frames) an afterglow 
in $\sim27\%$ of the cases. For time-to-observation of 
bursts $\geq500 {\rm ~s}$ and for Galactic reddening
$<0.5$ the UVOT pipeline detects 
in a {\em single} exposure an afterglow 
$\sim22\%$ of the time. If these samples of early
and late observed bursts are subdivided into long
($T_{90}>2 {\rm ~s}$) and short ($T_{90}\leq2 {\rm ~s}$) 
bursts, then the detection rate for a single exposure
is $\sim27\%$ for both long and short for the early
observed bursts, while it is $\sim29\%$ for the long
and $\sim12\%$ for the short late observed bursts.
These values will increase as future versions of
the catalog use optimal coaddition (see Section 6).
Initial work indicates that for bursts observed within 
$500 {\rm ~s}$ and for Galactic reddening
$<0.5$ the UVOT pipeline detects an afterglow 
in $\sim40\%$ of the cases. Many of the remaining 
$\sim60\%$ of ``dark" bursts can be explained by circumburst extinction, 
high redshift Lyman-$\alpha$ blanketing and absorption, and
suppression of the reverse shock 
\citep[cf.][]{RPWA06b,FJU01,GPJ98,HJP98}.

We have fit power law models to the light curves of 
bursts that are well-sampled in the UVOT observations.
The definition of ``well-sampled'' here is that for a 
single band, an afterglow must be detected in at least 
two independent images taken between 300\,s and 
$1\times 10^{5}\,s$ after the burst. A total of 
42 bursts are considered ``well-sampled."

Often the first 
few points on a light curve, up to several hundred 
seconds after a burst, are not consistent with a 
single power law fit that describes the rest of the 
light curve \citep[i.e. GRBs 050730, 050820A, 
060124, 060206, \& 060614; cf.][]{OS08}. 
In those cases, the early 
points have been omitted from the power law fit. The 
remaining points were fit with a 
function consisting of a single power law and a 
constant offset. The constant was included to account 
for any remaining residual in the sky subtraction, or 
to account for possible host galaxy contribution to 
the flux. The best fit parameters were determined by 
minimizing the $\chi^{2}$ value of the fit to the 
coincidence loss corrected linear count rates and 
their errors. The values of the fit parameters are 
given in the {\em Swift} UVOT Burst Catalog. Light 
curves and fits are shown in Figure\,9, for the 
well-sampled bursts in the bands for which there is 
the largest number of detections of a given burst 
afterglow. In the case of GRB\,060218, a power-law is 
not a satisfactory description of the light curve, so 
its light curve is shown without a fit. 
Figure~\ref{fig-061021} shows the light curves in all 
seven UVOT filters for GRB\,061007. The lower right 
panel of Figure~\ref{fig-061021}
shows all of the light curves for GRB\,061007 scaled 
to match the $v$ band light curve fit at $2000 
{\rm ~s}$. 

We note that the light curves have
been plotted on a log-linear scale as opposed
to the traditional log-log scale. This is done 
for diagnostic reasons in order to examine 
if the light curves approach 
zero. If the light curve approaches zero
then there is no host galaxy contributing
to the overall results. If the light curve
is above zero then there is host galaxy
contribution. If the light curve was below 
zero then a problem was identified and
fixed. 

Up to this point, we have only illustrated
light curves on the basis of the image database. 
Figure~\ref{fig-event} demonstrates the light curve for
GRB 060607A as generated from a version of the event 
database. The event database provides the capability
to probe the very early development of the afterglow. A
discussion of the early afterglow features is presented
elsewhere \citep{OS08}.

From the light curves a study of the interrelationship
between bursts can be made. The majority of the 
well-sampled light curves are
fit by a single power-law after several hundred 
seconds. Figure~\ref{fig-alpha} illustrates the
distribution of temporal slopes across all filters and
bursts while Figure~\ref{fig-alphafilters} shows 
the distribution of temporal slopes as a function of
filter. The temporal slope ranges from $-0.09$ to $-3$.
The average temporal slope for the entire sample
is $\alpha = 0.96$ (the dispersion about the mean, $\sigma = 0.48$) which is consistent with
other published results \citep{KDA07}. For the individual
filters the average $\alpha$ is 1.30 $(\sigma = 0.43)$, 
1.31 $(\sigma = 0.41)$, 0.96 $(\sigma = 0.33)$, 
0.86 $(\sigma = 0.38)$, 1.05 $(\sigma = 0.42)$, 
1.00 $(\sigma = 0.63)$, and 0.83 $(\sigma = 0.36)$ 
for the uvw2, uvm2, uvw1, $u$, $b$, $v$, and $white$
filters, respectively. A Kolmogorov-Smirnov (KS) test
between the UV and optical data sets (excluding 
the $white$ filter) indicates that the probability
of the two data sets being different is 0.087.

By using the
temporal slopes, we have calculated the magnitude
of the light curves at $2000 {\rm ~s}$ in each
available UVOT band. The average $v$-band magnitude
at $2000 {\rm ~s}$ is 18.02 ($\sigma = 1.59$).
Figure~\ref{fig-colors} reveals the color-color
relationship between the uvw1, $b$, and $v$ colors.
Typical values for $b-v$ are $\sim0.5$ with a small 
amount of scatter. On the other hand, values for
${\rm uvw1}-v$ are $\sim0$ with a large degree of
scatter. 

Figures~\ref{fig14}~\&~\ref{fig15} provide
a comparison of the X-ray flux at $11 {\rm ~hours}$ 
($F_{X,11}$) in the $0.3-10 {\rm ~keV}$ band, and the prompt $\gamma$-ray 
fluence ($S_\gamma$) in the $15-150 {\rm ~keV}$ band 
to the optical flux at $2000 {\rm ~s}$ 
($f_{2000}$)\footnote{The X-ray 
fluxes are provided by the {\em Swift} XRT team and 
will be published in an upcoming paper \citep{BDN08}.
The prompt $\gamma$-ray fluences are
found in \citet{ST2007}.}. Evident from the figures 
is a general trend of 
X-ray flux or $\gamma$-ray fluence to correspond to 
optical brightness. Using the Spearman rank
correlation, the data are strongly correlated
($p = 8.8\times10^{-4}$) and marginally correlated
($p = 0.0184$) for $F_{X,11}$ and $S_\gamma$,
respectively. An X-ray-to-optical correlation has
been suggested previously \citep{RPWA06b,RE05,JP04,DM03}.
We note that none of these values have been 
corrected for redshift. This is outside the scope of
this paper and left for future work.

We also provide the redshift distribution of the burst sample
(see Figure~\ref{fig-z}). The redshifts were not
determined by the UVOT but rather by spectroscopic measurements
with ground based instruments. From the distribution
it can be seen that the UVOT is sampling bursts across
the entire redshift range up to the UVOT redshift limit
of $z\sim5.1$. 

\section{Future Work}
This catalog is useful for examining the relationship
between optically detected and undetected bursts. 
Future versions of the database will incorporate various enhancements
to the current version. Some of these enhancements include:
using filter dependent region files, more fully automating quality checks,
adding functionality to the FTOOLs {\tt uvotsource}, and
optimal coaddition of images.

For the current version of the database, composite region files
were used. Future versions would use filter dependent (non-composite)
region files for each burst. There are several cases where a dense population
of sources in the $v$ filter make for a complicated background region file, while
the field is not at all complicated in the UV filters. A better background
estimate can be determined if separate region files were used.

Future databases may be able to reclaim some exposures flagged as ``poor quality"
images. Our method for finding images contaminated by charge trails or
diffraction spikes is to sum all images in a given filter and observation
sequence, visually inspect them, and manually flag
contaminated images. This works well for small apertures, but
future databases will require photometry for each object in a wide range
of apertures, which would potentially require a different set of quality
files for each aperture size. A better way to handle this problem would be
to identify bright stars in each field and determine whether the spacecraft
roll angle would align any of them in the CCD readout direction so as to affect
a given aperture at the GRB location. This eliminates the need for manual 
checking of a large number of images. By automating the process only the individual
images that are contaminated would be flagged, as opposed to flagging entire observation
sequences of images. Future databases would also check the photometric stability using
field stars for each image.

Several improvements to the UVOT tools used in generating the 
database have been proposed. Currently, {\tt uvotsource}
is not capable of performing sigma clipping of background pixels. Adding
this functionality would improve the background computations. Since
{\tt uvotsource} uses a Gaussian
model of the background, it has difficulty in estimating images with extremely low
background counts as we often see in UV images. Adding a Poisson or 
Binomial background model would greatly improve our
results in some cases.

In this database version, no coaddition of the individual images was attempted.
Future work will include optimal coaddition of the data using
the method proposed by \citet{MA08}. Using this method
will identify more detections from this and future
databases.

\acknowledgments
We gratefully acknowledge the contributions from members of the 
{\em Swift} team at
the Pennsylvania State University (PSU), University College 
London/Mullard Space Science Laboratory
(MSSL), NASA/Goddard Space Flight Center, 
and
our subcontractors, who helped make this instrument possible.
This work is sponsored at PSU 
by NASA contract NAS5-00136 and at MSSL by funding from the
Science and Technology Facilities Council (STFC). 

\facility{Swift(UVOT)}

\clearpage

\begin{deluxetable}{ccccc}
\tablecolumns{5}
\tabletypesize{\small}
\tablecaption{{\em Swift}/UVOT Filter Characteristics\tablenotemark{a}
\label{tab5}}
\tablewidth{0pt}
\tablehead{
  \colhead{Filter} &
  \colhead{$\lambda_c$} &
  \colhead{FWHM} &
  \colhead{$m_z$} &
  \colhead{$f_\lambda$}\\
  \colhead{} &
  \colhead{(\AA)} &
  \colhead{(\AA)} &
  \colhead{(mag)} &
  \colhead{($10^{-16} {\rm ~ergs\,cm^{-2}\,s^{-1}\,\AA^{-1}}$)} 
}
\startdata
uvw2    & 1928 &  657 & $17.35\pm0.04$ & $6.20\pm0.10$ \\
uvm2    & 2246 &  493 & $16.82\pm0.03$ & $8.50\pm0.06$ \\
uvw1    & 2600 &  693 & $17.49\pm0.03$ & $4.00\pm0.10$ \\
$u$     & 3465 &  785 & $18.34\pm0.02$ & $1.63\pm0.02$ \\
$b$     & 4392 &  975 & $19.11\pm0.02$ & $1.47\pm0.01$ \\
$v$     & 5468 &  769 & $17.89\pm0.01$ & $2.61\pm0.01$ \\
$white$ & 3850 & 2600 & $20.29\pm0.04$ & $0.37\pm0.05$ \\
\enddata
\tablenotetext{a}{$\lambda_c$ is the central wavelength, $m_z$
is the zero point, and $f_\lambda$ is the flux density 
conversion factor for the filters. All values
are from \citet{PTS2007}, except for $\lambda_c$ and FWHM for the 
$white$ filter, which are from \citet{RPWA2005}.}
\end{deluxetable}

\begin{deluxetable}{lrrcl}
\tablecolumns{5}
\tabletypesize{\small}
\tablecaption{UVOT Observed GRBs Not In the Databases\label{tab6}}
\tablewidth{0pt}
\tablehead{
  \colhead{OBJECT} &
  \colhead{${\rm RA_{J2000}}$} &
  \colhead{${\rm DEC_{J2000}}$} &
  \colhead{ERROR} &
  \colhead{Comments}\\
  \colhead{} &
  \colhead{} &
  \colhead{} &
  \colhead{(arcsec)} &
  \colhead{}
}
\startdata
GRB050215A & 348.382 &    49.322 & 240 & {\em Swift} GRB\\
GRB050922A & 271.154 & $-32.024$ & 210 & INTEGRAL GRB\\
GRB050925  & 303.476 &    34.332 & 120 & Possible {\em Swift} SGR\\
GRB051114  & 226.266 &    60.156 &  90 & Untriggered {\em Swift} GRB\\
GRB060102  & 328.834 &  $-1.838$ & 168 & {\em Swift} GRB\\
GRB060114  & 195.277 &  $-4.748$ & 120 & INTEGRAL GRB\\
GRB060130  & 229.224 & $-36.912$ & 120 & INTEGRAL GRB\\
GRB060728  &  16.646 & $-41.390$ & 180 & Possible {\em Swift} GRB\\
GRB061218  & 149.238 & $-35.221$ & 240 & Weak, short {\em Swift} GRB\\
\enddata
\end{deluxetable}

\begin{deluxetable}{lcrcccr}
\tablecolumns{7}
\tabletypesize{\small}
\tablecaption{Selected Sample from the {\em Swift}/UVOT Image Mode Burst Database\label{tab1}}
\tablewidth{0pt}
\tablehead{
  \colhead{OBJECT} &
  \colhead{TSTART} &
  \colhead{EXPOSURE} &
  \colhead{FILTER} &
  \colhead{MAG} &
  \colhead{MAG\_ERR} &
  \colhead{SNR}\\
  \colhead{} &
  \colhead{} &
  \colhead{(seconds)} &
  \colhead{} &
  \colhead{} &
  \colhead{} &
  \colhead{}
}
\startdata
GRB050525 & 138672238.421 & 9.679 & V & 13.97 & 0.06 & 16.3\\ 
GRB050525 & 138672397.228 & 9.613 & B & 15.17 & 0.06 & 16.0\\ 
GRB050525 & 138677707.262 & 885.569 & UVM2 & 17.75 & 0.06 & 17.9\\ 
GRB050525 & 138700740.978 & 190.932 & U & 19.28 & 0.17 & 6.4\\ 
\enddata
\tablecomments{Table~\ref{tab1} is published in its entirety in the
electronic edition of the Astrophysical Journal. A portion is 
shown here for guidance regarding its form and content.}
\end{deluxetable}

\begin{deluxetable}{lcrcc}
\tablecolumns{5}
\tabletypesize{\small}
\tablecaption{Selected Sample from the {\em Swift}/UVOT Event Mode Burst Database\label{tab7}}
\tablewidth{0pt}
\tablehead{
  \colhead{OBJECT} &
  \colhead{TSTART} &
  \colhead{EXPOSURE} &
  \colhead{FILTER} &
  \colhead{MAG}\\
  \colhead{} &
  \colhead{} &
  \colhead{(seconds)} &
  \colhead{} &
  \colhead{}
}
\startdata
GRB060607a & 1.71350008.193 & 9.993 & white & 17.23\\ 
GRB060607a & 1.71350018.193 & 10.000 & white & 17.06\\ 
GRB060607a & 1.71350028.193 & 10.000 & white & 16.74\\ 
GRB060607a & 1.71350038.193 & 10.000 & white & 16.38\\ 
\enddata
\tablecomments{Table~\ref{tab7} is published in its entirety in the
electronic edition of the Astrophysical Journal. A portion is 
shown here for guidance regarding its form and content.}
\end{deluxetable}

\begin{deluxetable}{lcrccrc}
\tablecolumns{7}
\tabletypesize{\small}
\tablecaption{Selected Sample from the {\em Swift}/UVOT Burst Catalog\label{tab2}}
\tablewidth{0pt}
\tablehead{
  \colhead{OBJECT} &
  \colhead{RA\_BEST} &
  \colhead{DEC\_BEST} &
  \colhead{POS\_ERR} &
  \colhead{DISC\_BY} &
  \colhead{T90} &
  \colhead{TRIGTIME}
}
\startdata
GRB050730 & 212.071208 &  $-3.771917$ & 0.50 & 0 & 156.5 & 144446303.2\\ 
GRB050801 & 204.145833 & $-21.928055$ & 0.50 & 0 &  19.4 & 144613682.1\\ 
GRB050802 & 219.274250 &  27.786840 & 0.50 & 0 &  19.0 & 144670082.1\\ 
GRB050803 & 350.657958 &   5.785833 & 1.40 & 0 &  87.9 & 144789240.3\\ 
\enddata
\tablecomments{Table~\ref{tab2} is published in its entirety in the
electronic edition of the Astrophysical Journal. A portion is 
shown here for guidance regarding its form and content.}
\end{deluxetable}

\begin{deluxetable}{ccl}
\tablecolumns{3}
\tabletypesize{\small}
\tablecaption{{\em Swift}/UVOT Image Mode Burst Database Format\label{tab3}}
\tablewidth{0pt}
\tablehead{
  \colhead{Column} &
  \colhead{Format} &
  \colhead{Description}}
\startdata
1   & A10   & Object ID $GRByymmssX$\\
2   & F10.6 & Right ascension (J2000.0) in decimal degrees\\
3   & F10.6 & Declination (J2000.0) in decimal degrees\\
4   & F6.2  & Positional uncertainty in arcseconds\\
5   & F11.1 & Time of burst trigger in mission elapsed time (MET)\\
6   & A23   & Time of burst trigger in Universal time\\
7   & F13.3 & Time of middle of exposure in MET\\
8   & F13.3 & Time of start of exposure in MET\\
9  & F13.3 & Time of end of exposure in MET\\
10  & F8.3  & Total exposure time in seconds with all correction applied\\
11  & F8.3  & Total elapsed time in seconds\\
12  & A5    & UVOT filter\\
13  & I1    & Binning factor\\
14  & F4.1  & Source aperture radius in arcseconds\\
15  & F7.2  & Area of the source region (SR) in square arcseconds\\
16  & F8.2  & Area of the background region (BR) in square arcseconds\\
17  & F5.3  & Image plate scale in arcseconds per pixel\\
18  & F9.2  & Total number of counts in SR\\
19  & F9.2  & Error in the total number of counts in SR\\
20  & F9.2  & Total number of counts in BR\\
21  & F9.2  & Error in the total number of counts in BR\\
22  & F9.2  & Total number of counts in standard SR\\
23  & F9.2  & Error in the total number of counts in standard SR\\
24  & E11.4 & Count rate in SR in counts per second\\
25  & E11.4 & Error in the SR count rate\\
26  & E11.4 & Count rate in BR per square arcseconds\\
27  & E11.4 & Error in the BR count rate per square arcseconds\\
28  & E11.4 & Global background rate in counts per second per\\
    &       & square arcsecond\\
29  & E11.4 & Area of global background region in square arcseconds\\
30  & E11.4 & Count rate in coincidence loss region (CLR)\\
31  & E11.4 & Error in the count rate of the CLR\\
32  & F5.3  & Coincidence loss correction factor (CLCF) for CLR\\
33  & F5.3  & Error in the CLR CLCF\\
34  & F5.3  & CLCF for BR\\
35  & F5.3  & Error in the BR CLCF\\
36  & F9.2  & Coincidence loss corrected (CLC) counts in SR\\
37  & F9.2  & Error in the SR CLC counts\\
38  & F9.2  & CLC counts in BR\\
39  & F9.2  & Error in the BR CLC counts\\
40  & E11.4 & CLC raw count rate in SR\\
41  & E11.4 & Error in the SR CLC raw count rate\\
42  & E11.4 & CLC background surface count rate in counts per second per\\
    &       & square arcsecond\\
43  & E11.4 & Error in the CLC background surface count rate\\
44  & E11.4 & Coincidence corrected net count rate\\
45  & E11.4 & Error in the coincidence corrected net count rate\\
46  & F5.3  & Aperture correction between a $3\arcsec$ to $5\arcsec$ aperture\\
47  & F5.3  & Aperture factor used to compute the count rate error\\
48  & E11.4 & CLC count rate use to calculate flux and magnitude\\
49  & E11.4 & Error on the CLC count rate use to calculate flux and magnitude\\
50  & F5.2  & Magnitude/upper-limit of the source (99.00 for upper-limit)\\
51  & F5.2  & $1\sigma$ error in magnitude (9.90E+01 for upper-limits)\\
52  & F6.1  & Signal-to-noise ratio calculated from columns 41 \& 43\\
53  & F5.2  & Magnitude of the background per square arcseconds\\
54  & F5.2  & $1\sigma$ error in background in magnitude per square arcseconds\\
55  & F5.2  & ``N"-sigma limiting magnitude\\
56  & F3.1  & Chosen parameter for ``N"\\
57  & F5.2  & Magnitude at which the count rate is one count per CCD frame\\
58  & E10.3 & Flux density in ${\rm ~erg ~s^{-1} ~cm^{-2} ~\AA^{-1}}$\\
59  & E10.3 & $1\sigma$ error in flux density in ${\rm ~erg ~s^{-1} ~cm^{-2} ~\AA^{-1}}$\\
60  & E10.3 & Flux density of the sky in ${\rm ~erg ~s^{-1} ~cm^{-2} ~\AA^{-1} ~arcsec^{-2}}$\\
61  & E10.3 & $1\sigma$ error in sky flux density in ${\rm ~erg ~s^{-1} ~cm^{-2} ~\AA^{-1} ~arcsec^{-2}}$\\
62  & E10.3 & ``N"-sigma limiting flux density in ${\rm ~erg ~s^{-1} ~cm^{-2} ~\AA^{-1}}$\\
63  & E10.3 & Flux density in ${\rm ~erg ~s^{-1} ~cm^{-2} ~\AA^{-1}}$ at which the count rate is\\
    &       & one count per frame\\
64  & E10.3 & Flux density in ${\rm ~mJy}$\\
65  & E10.3 & $1\sigma$ error in flux density in ${\rm ~mJy}$\\
66  & E10.3 & Flux density of the sky in ${\rm ~mJy ~arcsec^{-2}}$\\
67  & E10.3 & $1\sigma$ error in sky flux density in ${\rm ~mJy ~arcsec^{-2}}$\\
68  & E10.3 & ``N"-sigma limiting flux density in ${\rm ~mJy}$\\
69  & E10.3 & Flux density in ${\rm ~mJy}$ at which the count rate is\\
    &       & one count per frame\\
70  & F10.6 & Right ascension (J2000.0) in decimal degrees of closest\\
    &       & non-GRB source\\
71  & F10.6 & Declination (J2000.0) in decimal degrees of closest\\
    &       & non-GRB source\\
72  & F6.2  & Magnitude in the given UVOT band of closest non-GRB source\\
73  & F10.6 & Centroided right ascension (J2000.0) in decimal degrees\\
74  & F10.6 & Centroided declination (J2000.0) in decimal degrees\\
75  & F5.1  & Centroided positional uncertainty in arcseconds\\
76  & A1    & Settling image flag (T/F)\\
77  & A1    & Aspect error flag (T/F)\\
78  & A1    & Charge trailing flag (T/F)\\
79  & A1    & Crowded field flag (T/F)\\
80  & A1    & Diffraction spike flag (T/F)\\
81  & A1    & Edge of field flag (T/F)\\
82  & A1    & In halo flag (T/F)\\
83  & A1    & Poor quality flag (T/F)\\
84  & A8    & Swift trigger number\\
85  & A12   & Name of source FITS extension\\
86  & A54   & Image name\\
\enddata
\end{deluxetable}

\begin{deluxetable}{ccl}
\tablecolumns{3}
\tabletypesize{\small}
\tablecaption{{\em Swift}/UVOT Event Mode Burst Database Format\label{tab8}}
\tablewidth{0pt}
\tablehead{
  \colhead{Column} &
  \colhead{Format} &
  \colhead{Description}}
\startdata
1   & A10   & Object ID $GRByymmssX$\\
2   & F10.6 & Right ascension (J2000.0) in decimal degrees\\
3   & F10.6 & Declination (J2000.0) in decimal degrees\\
4   & F6.2  & Positional uncertainty in arcseconds\\
5   & F13.3 & Time of burst trigger in MET\\
6   & A23   & Time of burst trigger in Universal time\\
7   & F13.3 & Time of middle of exposure in MET\\
8   & F13.3 & Time of start of exposure in MET\\
9   & F13.3 & Time of end of exposure in MET\\
10  & F8.3  & Total exposure time in seconds with all correction applied\\
11  & F8.3  & Total elapsed time in seconds\\
12  & A5    & UVOT filter\\
13  & I1    & Binning factor\\
14  & F4.1  & Source aperture radius in arcseconds\\
15  & F7.2  & Area of the SR in square arcseconds\\
16  & F10.2 & Area of the BR in square arcseconds\\
17  & F5.3  & Image plate scale in arcseconds per pixel\\
18  & F6.3  & CLCF for CLR\\
19  & F5.3  & Error in the CLR CLCF\\
20  & F5.3  & CLCF for BR\\
21  & F5.3  & Error in the BR CLCF\\
22  & E11.4 & CLC raw count rate in SR\\
23  & E11.4 & Error in the SR CLC raw count rate\\
24  & E11.4 & CLC background surface count rate in counts per second per\\
    &       & square arcsecond\\
25  & E11.4 & Error in the CLC background surface count rate\\
26  & E11.4 & Coincidence corrected net count rate\\
27  & E11.4 & Error in the coincidence corrected net count rate\\
28  & F5.3  & Aperture correction between a $3\arcsec$ to $5\arcsec$ aperture\\
29  & E11.4 & CLC count rate use to calculate flux and magnitude\\
30  & E11.4 & Error on the CLC count rate use to calculate flux and magnitude\\
31  & F5.2  & Magnitude/upper-limit of the source (99.00 for upper-limit)\\
32  & F5.2  & $1\sigma$ error in magnitude (9.90E+01 for upper-limits)\\
33  & E10.3 & Flux density in ${\rm ~erg ~s^{-1} ~cm^{-2} ~\AA^{-1}}$\\
34  & E10.3 & $1\sigma$ error in flux density in ${\rm ~erg ~s^{-1} ~cm^{-2} ~\AA^{-1}}$\\
35  & F10.6 & Centroided right ascension (J2000.0) in decimal degrees\\
36  & F10.6 & Centroided declination (J2000.0) in decimal degrees\\
37  & F5.1  & Centroided positional uncertainty in arcseconds\\
38  & A1    & Settling image flag (T/F)\\
39  & A1    & Aspect error flag (T/F)\\
40  & A8    & Swift trigger number\\
41  & A54   & Event list filename\\
\enddata
\end{deluxetable}

\begin{deluxetable}{ccl}
\tablecolumns{3}
\tabletypesize{\small}
\tablecaption{{\em Swift}/UVOT Burst Catalog Format\label{tab4}}
\tablewidth{0pt}
\tablehead{
  \colhead{Column} &
  \colhead{Format} &
  \colhead{Description}}
\startdata
1   & A10   & Object ID $GRByymmssX$\\
2   & F10.6 & Right ascension (J2000.0) in decimal degrees\\
3   & F10.6 & Declination (J2000.0) in decimal degrees\\
4   & F6.2  & Positional uncertainty in arcseconds\\
5   & A8    & Reference for the given position and error\\
6   & I1    & Discovery flag\\
7   & F5.1  & $T_{90}$ in seconds\\
8   & F11.1 & Time of burst trigger in mission elapsed time (MET)\\
9   & A23   & Time of burst trigger in Universal time\\
10  & E13.6 & Start time in seconds of 1st UVOT observation\\
11  & F6.2  & First reported magnitude (FRM) in the uvw2-filter\\
12  & E13.6 & Time since burst, in seconds, for the FRM in the uvw2-filter\\
13  & F6.2  & FRM in the uvm2-filter\\
14  & E13.6 & Time since burst, in seconds, for the FRM in the uvm2-filter\\
15  & F6.2  & FRM in the uvw1-filter\\
16  & E13.6 & Time since burst, in seconds, for the FRM in the uvw1-filter\\
17  & F6.2  & FRM in the $u$-filter\\
18  & E13.6 & Time since burst, in seconds, for the FRM in the $u$-filter\\
19  & F6.2  & FRM in the $b$-filter\\
20  & E13.6 & Time since burst, in seconds, for the FRM in the $b$-filter\\
21  & F6.2  & FRM in the $v$-filter\\
22  & E13.6 & Time since burst, in seconds, for the FRM in the $v$-filter\\
23  & F6.2  & FRM in the $white$-filter\\
24  & E13.6 & Time since burst, in seconds, for the FRM in the $white$-filter\\
25  & F6.2  & Peak reported magnitude (PRM) in the uvw2-filter\\
26  & E13.6 & Time since burst, in seconds, for the PRM in the uvw2-filter\\
27  & F6.2  & PRM in the uvm2-filter\\
28  & E13.6 & Time since burst, in seconds, for the PRM in the uvm2-filter\\
29  & F6.2  & PRM in the uvw1-filter\\
30  & E13.6 & Time since burst, in seconds, for the PRM in the uvw1-filter\\
31  & F6.2  & PRM in the $u$-filter\\
32  & E13.6 & Time since burst, in seconds, for the PRM in the $u$-filter\\
33  & F6.2  & PRM in the $b$-filter\\
34  & E13.6 & Time since burst, in seconds, for the PRM in the $b$-filter\\
35  & F6.2  & PRM in the $v$-filter\\
36  & E13.6 & Time since burst, in seconds, for the PRM in the $v$-filter\\
37  & F6.2  & PRM in the $white$-filter\\
38  & E13.6 & Time since burst, in seconds, for the PRM in the $white$-filter\\
39  & F6.2  & Temporal slope in the uvw2-filter\\
40  & F6.2  & Error in the temporal slope of the uvw2-filter\\
41  & E9.2  & Amplitude in the temporal slope of the uvw2-filter\\
42  & F6.2  & Magnitude in the uvw2-filter at $2000 {\rm ~s}$\\
43  & F6.2  & Temporal slope in the uvm2-filter\\
44  & F6.2  & Error in the temporal slope of the uvm2-filter\\
45  & E9.2  & Amplitude in the temporal slope of the uvm2-filter\\
46  & F6.2  & Magnitude in the uvm2-filter at $2000 {\rm ~s}$\\
47  & F6.2  & Temporal slope in the uvw1-filter\\
48  & F6.2  & Error in the temporal slope of the uvw1-filter\\
49  & E9.2  & Amplitude in the temporal slope of the uvw1-filter\\
50  & F6.2  & Magnitude in the uvw1-filter at $2000 {\rm ~s}$\\
51  & F6.2  & Temporal slope in the $u$-filter\\
52  & F6.2  & Error in the temporal slope of the $u$-filter\\
53  & E9.2  & Amplitude in the temporal slope of the $u$-filter\\
54  & F6.2  & Magnitude in the $u$-filter at $2000 {\rm ~s}$\\
55  & F6.2  & Temporal slope in the $b$-filter\\
56  & F6.2  & Error in the temporal slope of the $b$-filter\\
57  & E9.2  & Amplitude in the temporal slope of the $b$-filter\\
58  & F6.2  & Magnitude in the $b$-filter at $2000 {\rm ~s}$\\
59  & F6.2  & Temporal slope in the $v$-filter\\
60  & F6.2  & Error in the temporal slope of the $v$-filter\\
61  & E9.2  & Amplitude in the temporal slope of the $v$-filter\\
62  & F6.2  & Magnitude in the $v$-filter at $2000 {\rm ~s}$\\
63  & F6.2  & Temporal slope in the $white$-filter\\
64  & F6.2  & Error in the temporal slope of the $white$-filter\\
65  & E9.2  & Amplitude in the temporal slope of the $white$-filter\\
66  & F6.2  & Magnitude in the $white$-filter at $2000 {\rm ~s}$\\
67  & F6.2  & $b-v$ at $2000 {\rm ~s}$\\
68  & F6.2  & ${\rm uvw1}-v$ at $2000 {\rm ~s}$\\
69  & F5.3  & Galactic reddening\\
70  & E8.2  & Logarithm of Galactic HI column density (log $N_H$)\\
71  & F7.4  & Redshift\\
72  & I1    & Redshift type flag\\
73  & I4    & GCN Circular number where redshift was reported\\
74  & E9.2  & Prompt BAT fluence in ${\rm ~erg ~cm^{-2}}$ as reported by \citet{ST2007}\\
75  & F7.3  & Radio flux in ${\rm ~mJy}$ as reported in GCN\\
76  & I4    & GCN Circular number where radio flux was reported\\
77  & F6.2  & Frequency in GHz of radio detection as reported in GCN\\
78  & A1    & Flag indicating a detection in the $R-K$ band was reported in GCN Circulars\\
79  & A1    & Flag indicating a detection in a UVOT band\\
80  & A1    & Flag indicating a detection in the radio was reported in GCN Circulars\\
81  & A256  & Notes\\
\enddata
\end{deluxetable}

\clearpage

\begin{figure}
\figurenum{9a}
\epsscale{0.8}
\plotone{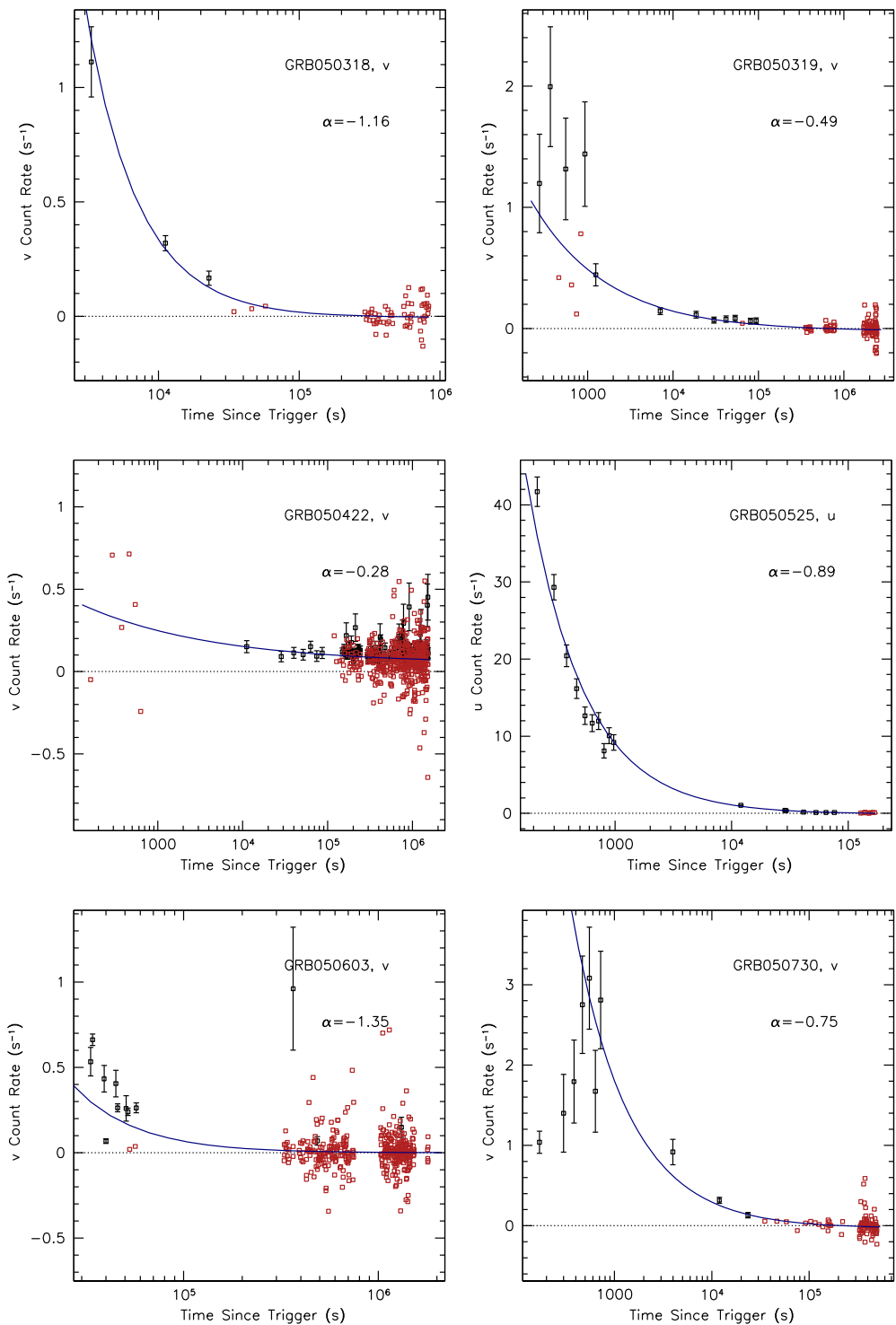}
\caption{Light curves of all ``well" sampled bursts in
one UVOT band (part 1 of 7). The name of the burst, the UVOT band
displayed, and the temporal slope are provided. Black
points are UVOT detections and red points are upper 
limits. [{\em See the electronic edition of
the Journal for a color version of this figure.}]}
\label{fig-alllcs}   
\end{figure}

\begin{figure}
\figurenum{9b}
\epsscale{0.9}
\plotone{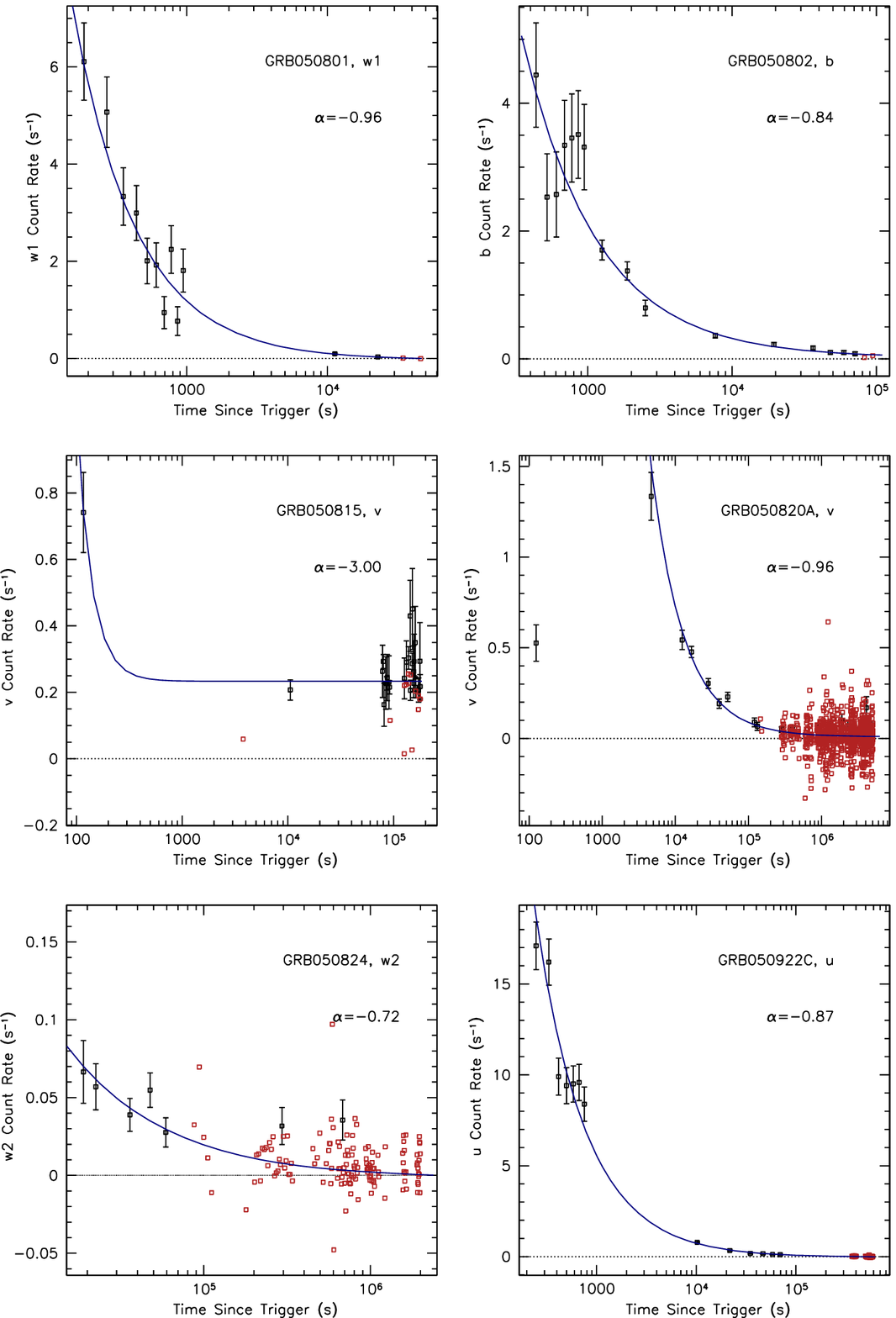}
\caption{Light curves of all ``well" sampled bursts in
one UVOT band (part 2 of 7).}
\end{figure}

\begin{figure}
\figurenum{9c}
\epsscale{0.9}
\plotone{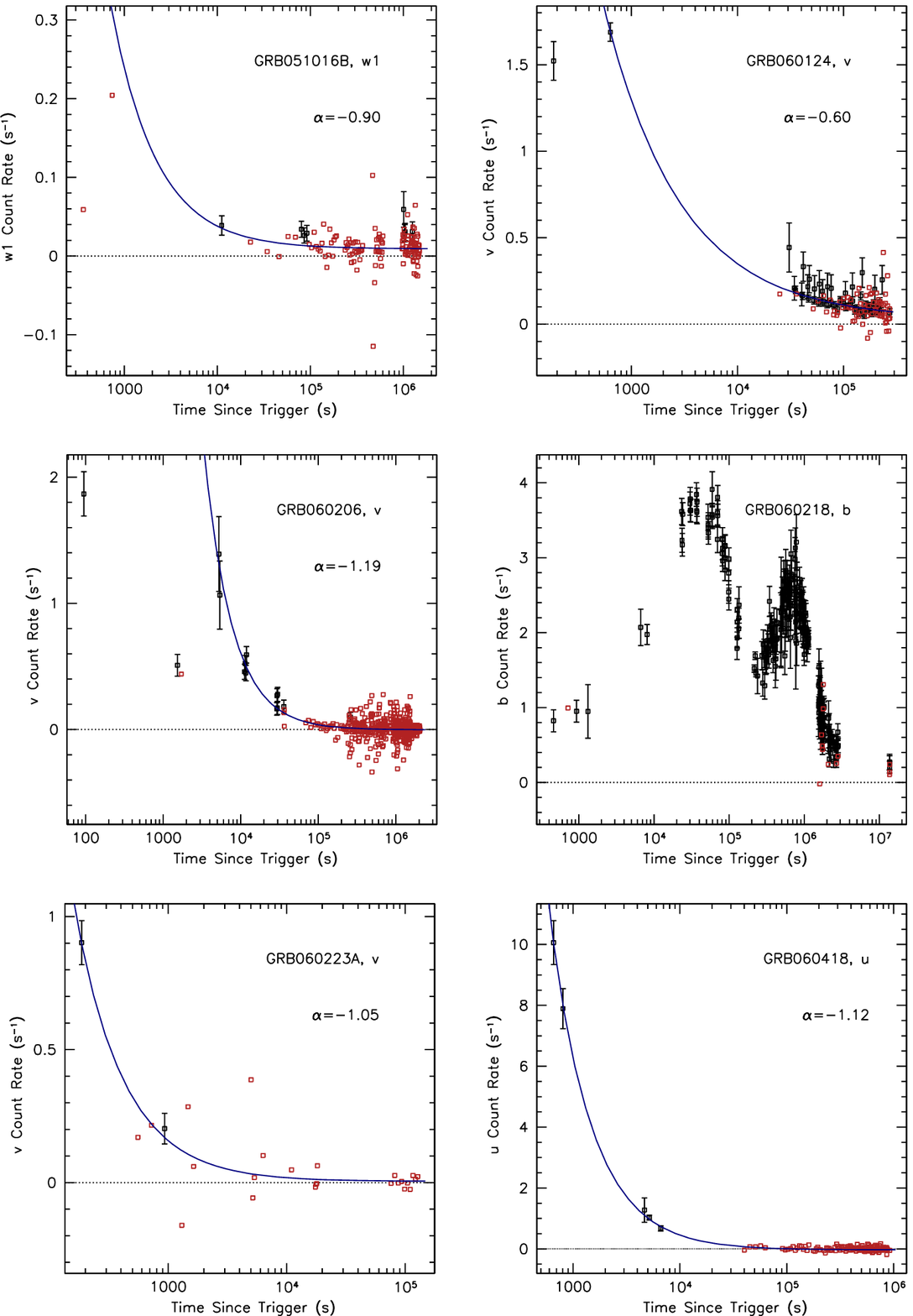}
\caption{Light curves of all ``well" sampled bursts in
one UVOT band (part 3 of 7).}
\end{figure}

\begin{figure}
\figurenum{9d}
\epsscale{0.9}
\plotone{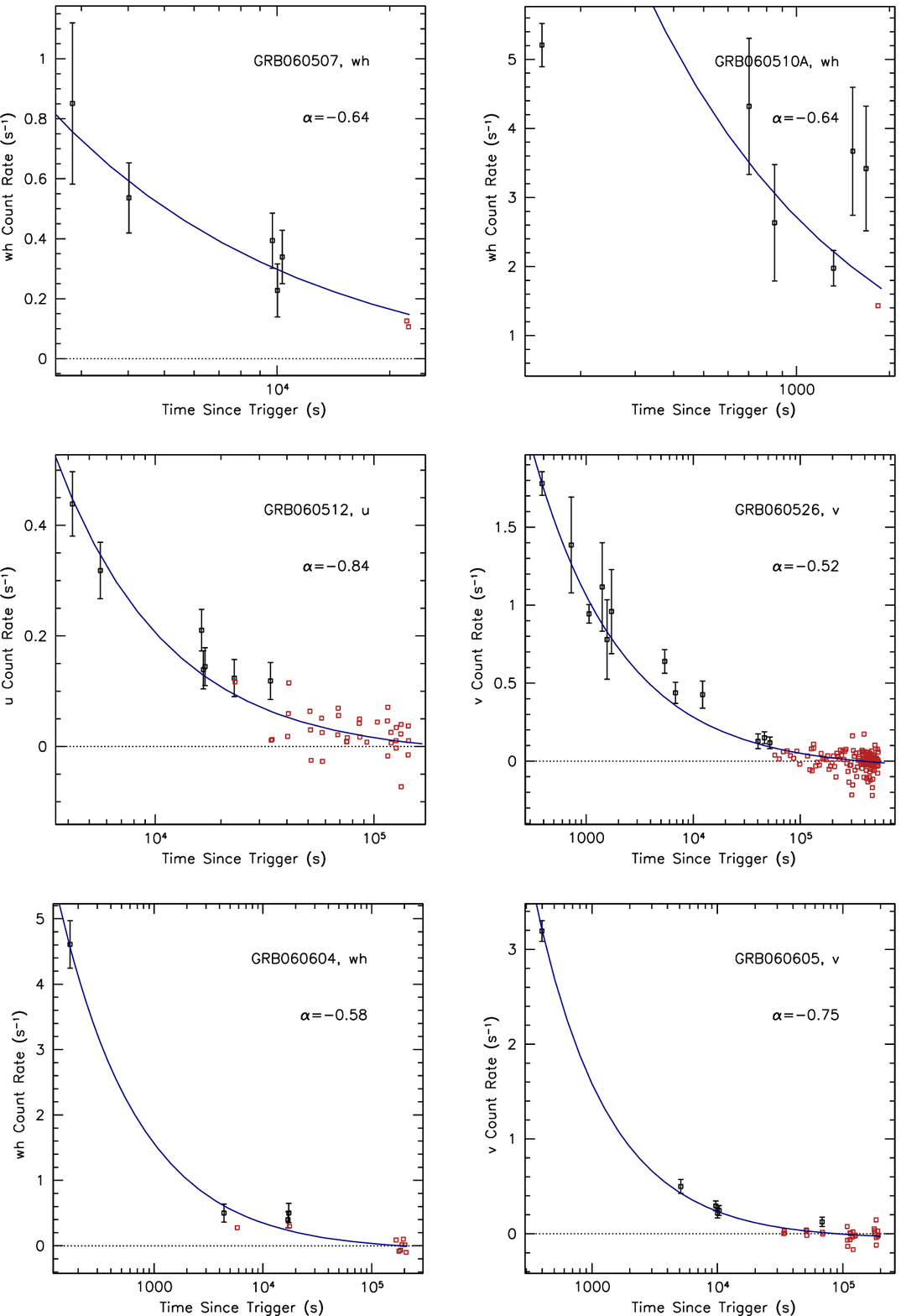}
\caption{Light curves of all ``well" sampled bursts in
one UVOT band (part 4 of 7).}
\end{figure}

\begin{figure}
\figurenum{9e}
\epsscale{0.9}
\plotone{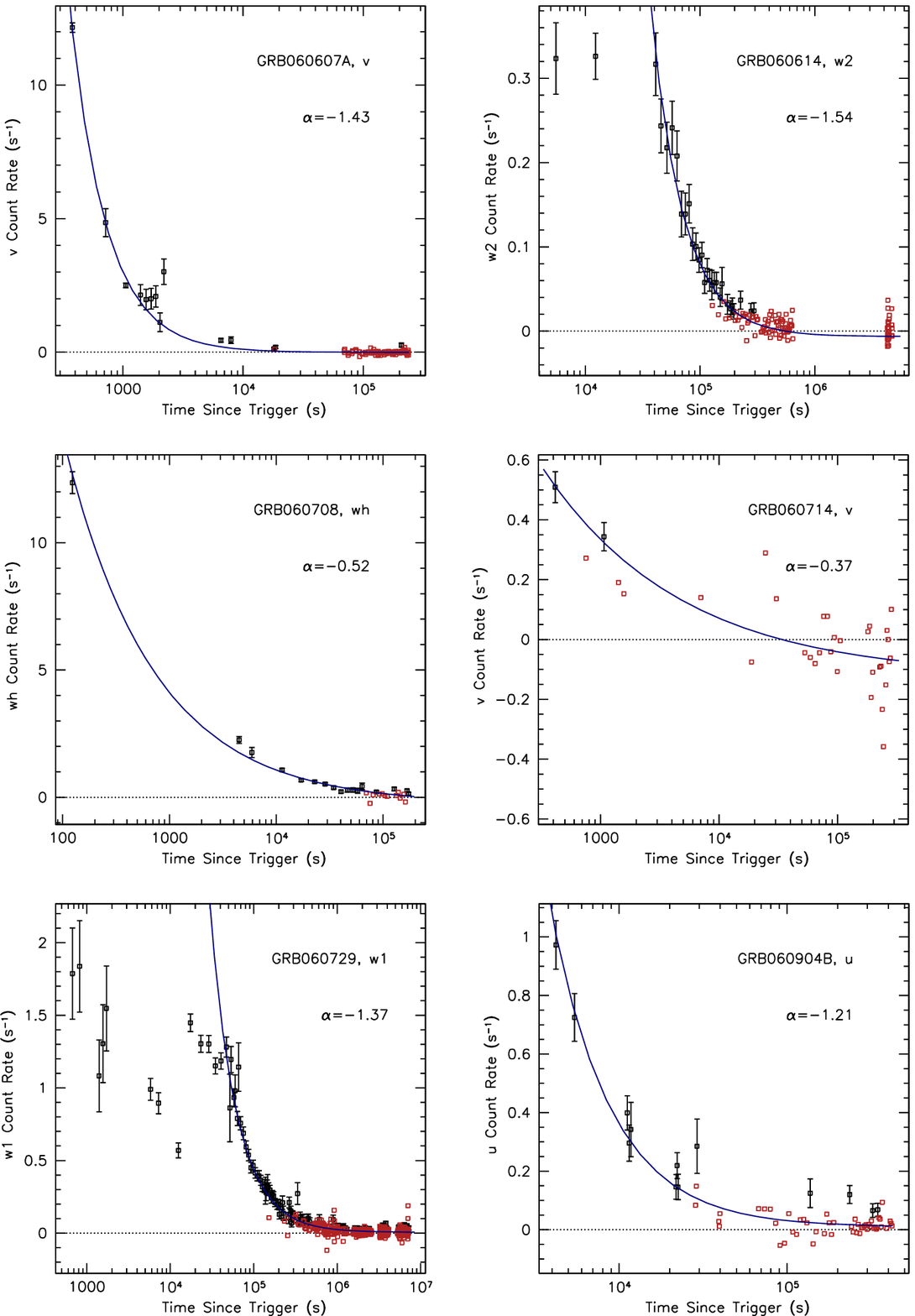}
\caption{Light curves of all ``well" sampled bursts in
one UVOT band (part 5 of 7).}
\end{figure}

\begin{figure}
\figurenum{9f}
\epsscale{0.9}
\plotone{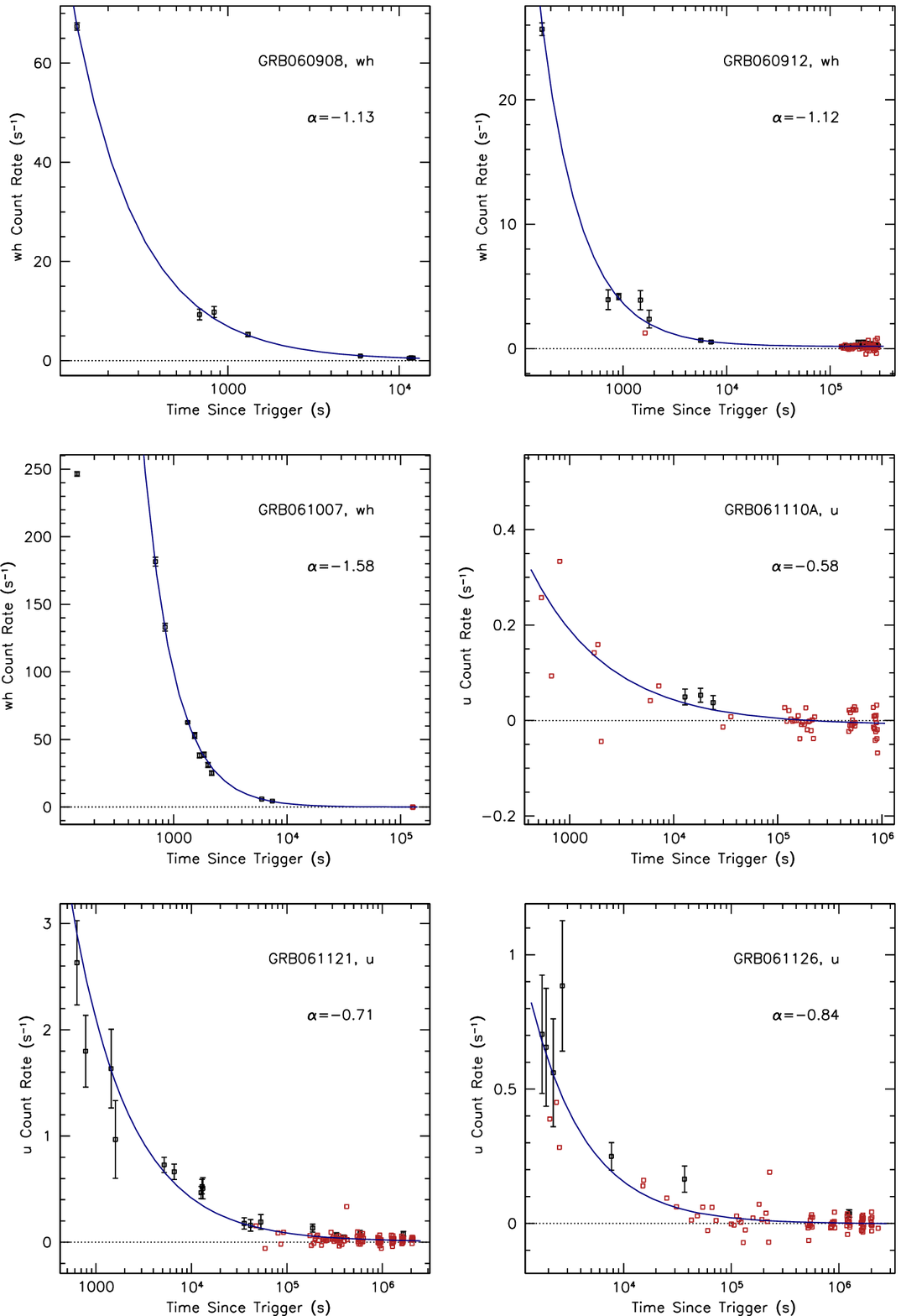}
\caption{Light curves of all ``well" sampled bursts in
one UVOT band (part 6 of 7).}
\end{figure}

\begin{figure}
\figurenum{9g}
\epsscale{0.9}
\plotone{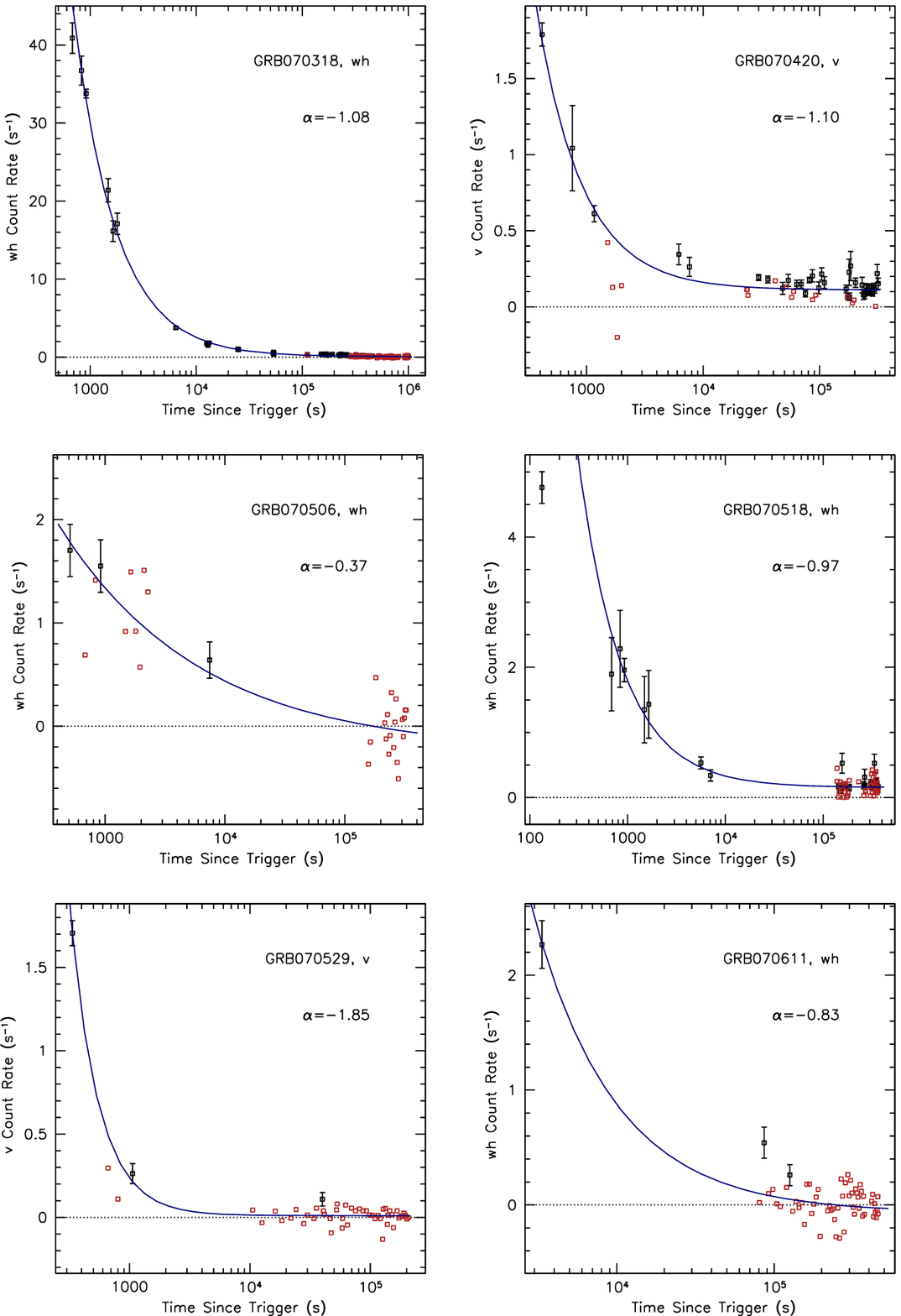}
\caption{Light curves of all ``well" sampled bursts in
one UVOT band (part 7 of 7).}
\end{figure}

\clearpage

\setcounter{figure}{9}
\begin{figure}
\epsscale{0.9}
\plotone{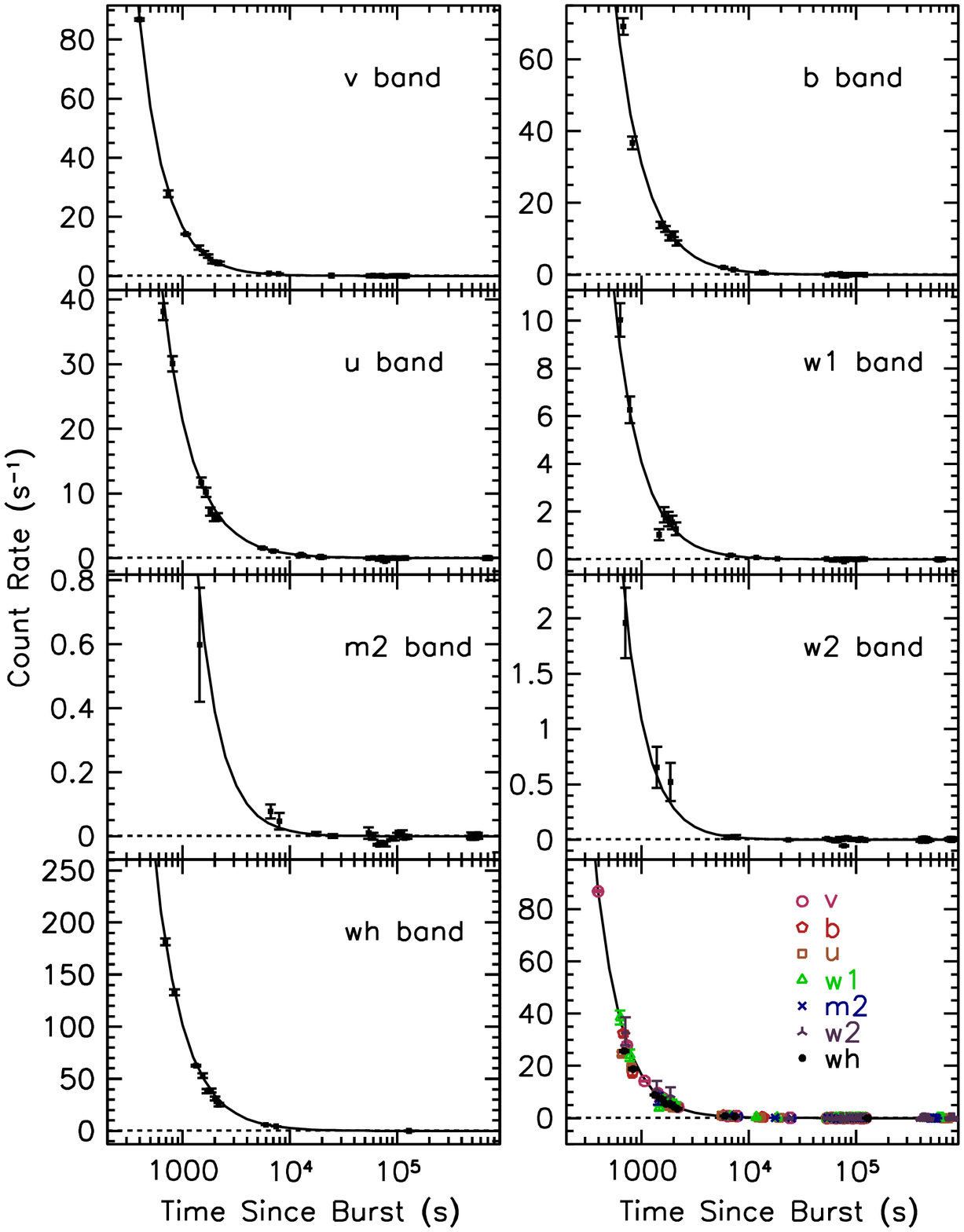}
\caption{Light curves in each UVOT band for GRB 061007. 
[{\em See the electronic edition of
the Journal for a color version of this figure.}]}
\label{fig-061021}   
\end{figure}

\begin{figure}
\epsscale{0.9}
\plotone{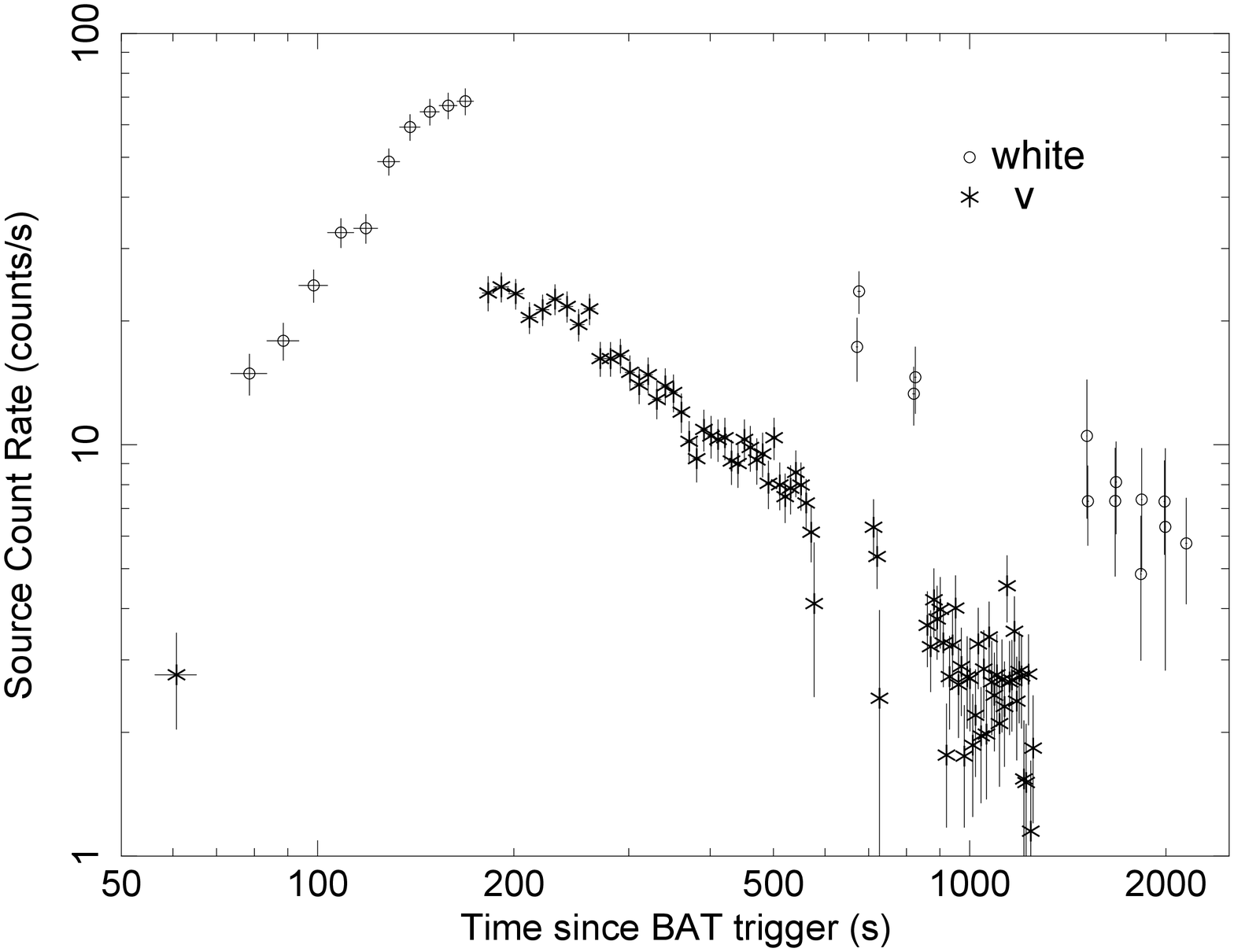}
\caption{Light curve for GRB 060607A produced
from a version of the event database. Typically, the taking of 
data in event mode has ceased by $\sim2000 {\rm ~s}$.
The count rate is in ${\rm counts ~s^{-1}}$. The
$v$-band data point at $\sim62 {\rm ~s}$ is 
produced while the spacecraft is settling on
the target. The UVOT detector voltage is still
changing during this period, therefore, the
count rate is not calibrated for this situation.}
\label{fig-event}   
\end{figure}

\begin{figure}
\epsscale{0.9}
\plotone{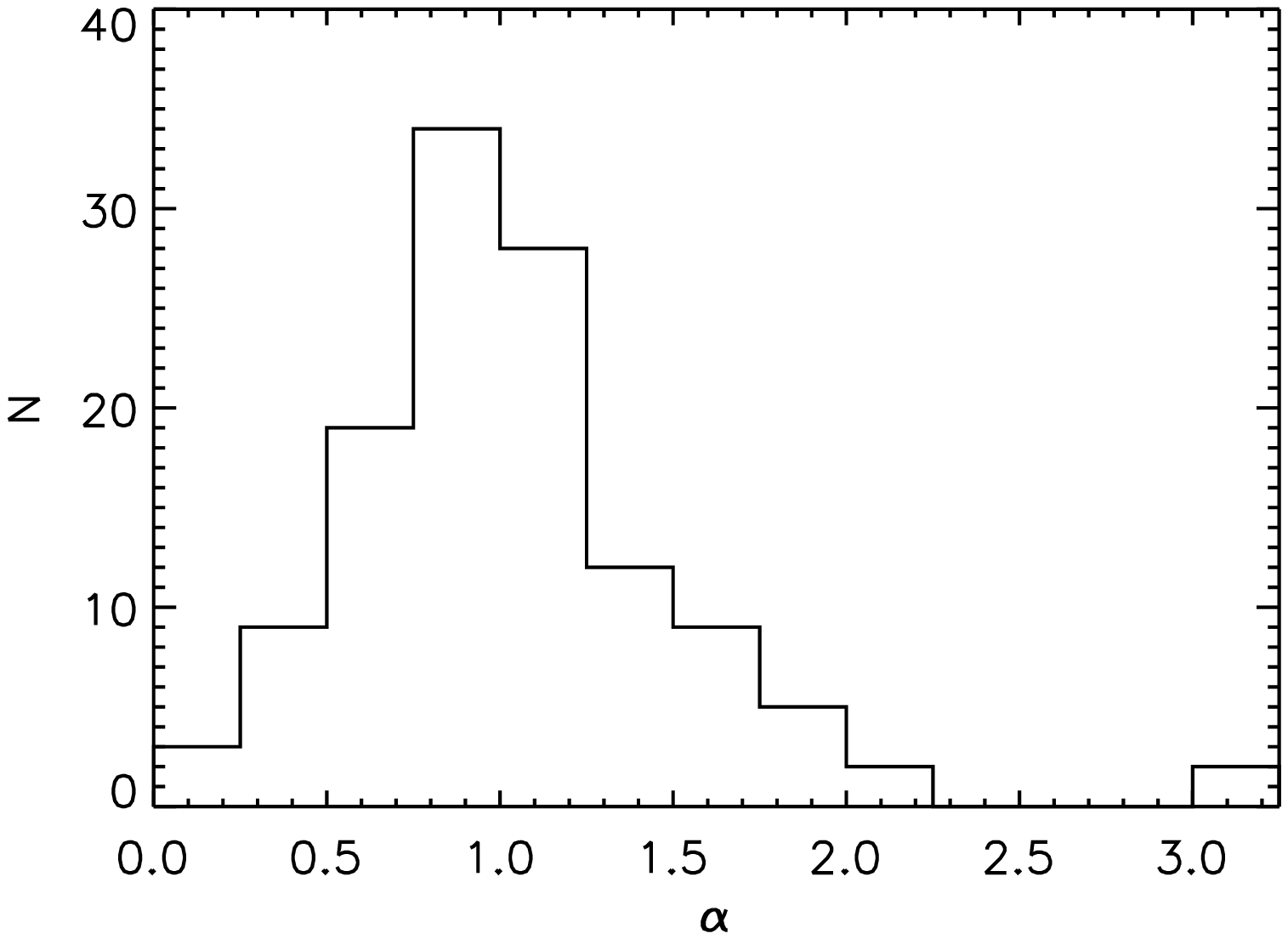}
\caption{The filter independent temporal slope 
distribution for the 42 ``well" sampled GRBs. 
The median temporal 
slope is $\alpha = 0.96 ~(\sigma = 0.48)$.}
\label{fig-alpha}   
\end{figure}

\begin{figure}
\epsscale{0.5}
\plotone{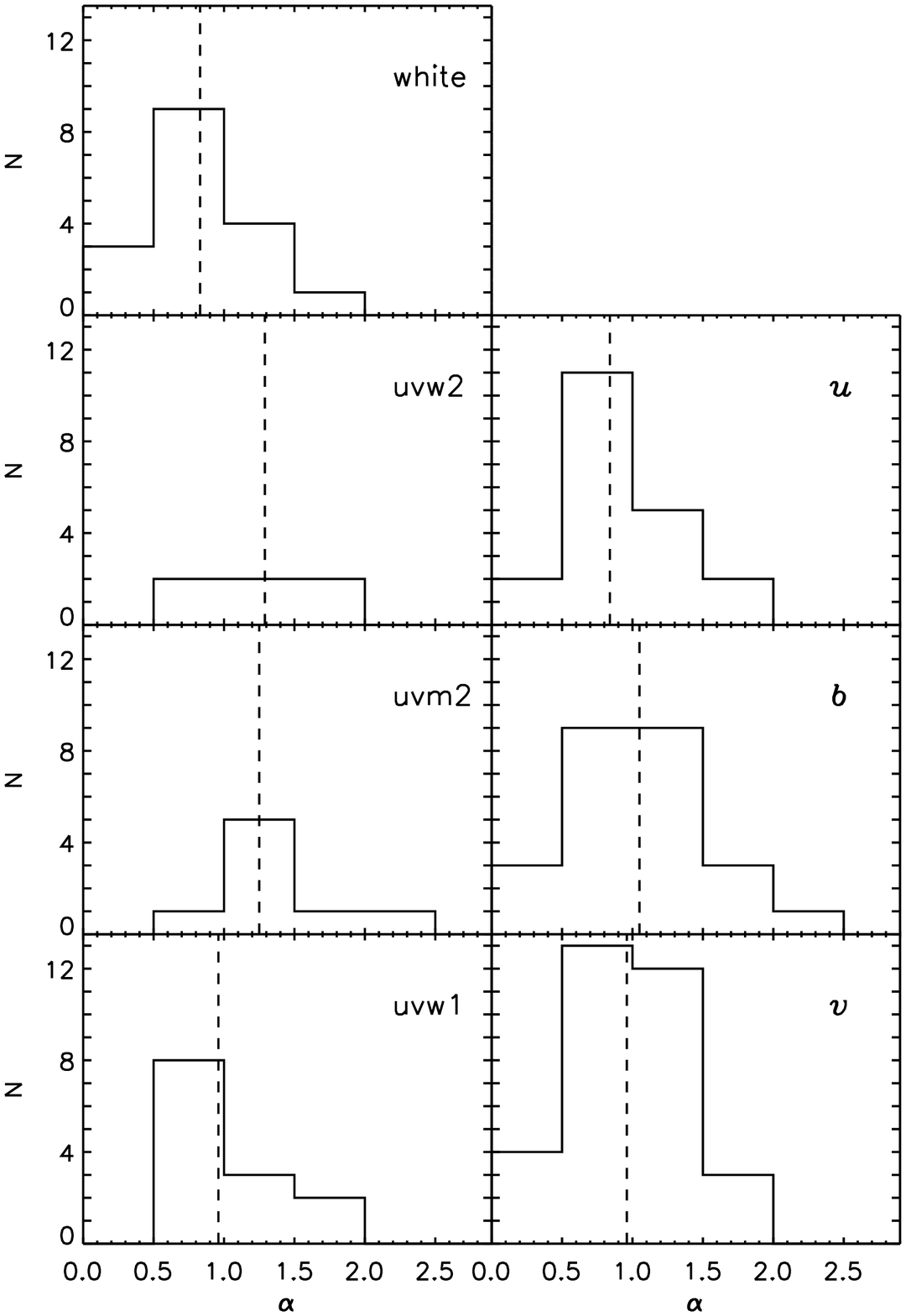}
\caption{Temporal slope distribution in different
UVOT filters. The median temporal slope 
is 1.30 $(\sigma = 0.43)$, 
1.31 $(\sigma = 0.41)$, 0.96 $(\sigma = 0.33)$, 
0.86 $(\sigma = 0.38)$, 1.05 $(\sigma = 0.42)$, 
1.00 $(\sigma = 0.63)$, and 0.83 $(\sigma = 0.36)$ 
for the uvw2, uvm2, uvw1, $u$, $b$, $v$, and $white$
filters, respectively.}
\label{fig-alphafilters}   
\end{figure}

\begin{figure}
\epsscale{0.9}
\plotone{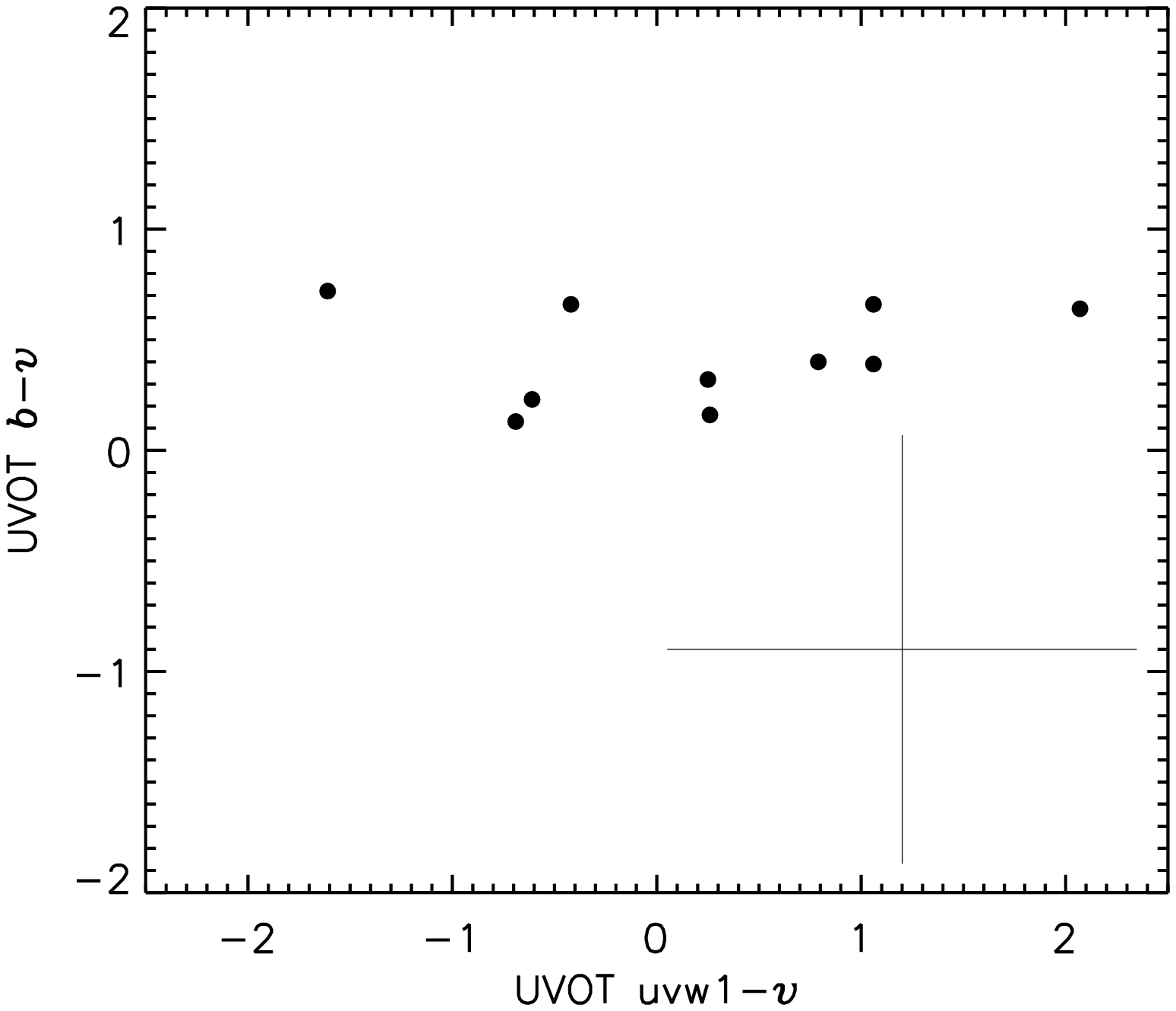}
\caption{Distribution of afterglow colors as 
measured by the UVOT for bursts in our sample. 
The colors are all taken from $2000 {\rm ~s}$
post burst trigger. 
The central wavelengths for the uvw1, $b$, and 
$v$ filters are 2600, 4392, and 5468 \AA, 
respectively. The error bar denotes the
median errors for an individual measurement. 
Only bursts with detections in
all three bands are represented.}
\label{fig-colors}   
\end{figure}

\begin{figure}
\epsscale{0.9}
\plotone{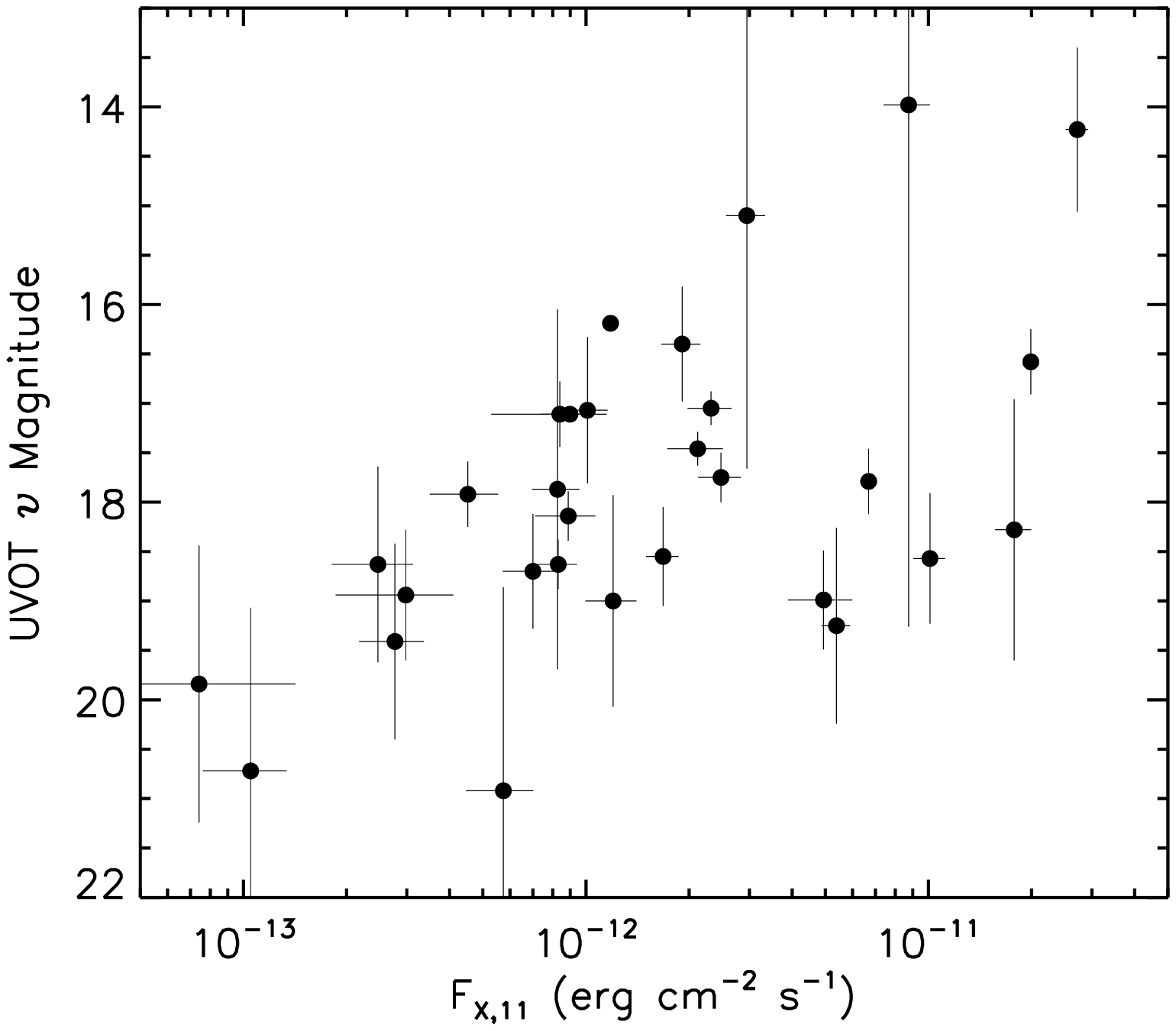}
\caption{Comparison of the XRT flux at $11 {\rm
~hours}$ ($F_{X,11}$) in the $0.3-10 {\rm ~keV}$ 
band versus the UVOT magnitude in the $v$-band
at $2000 {\rm ~s}$. Using the Spearman rank
correlation, the data are strongly correlated
($p = 8.8\times10^{-4}$).}
\label{fig14}   
\end{figure}

\begin{figure}
\epsscale{0.9}
\plotone{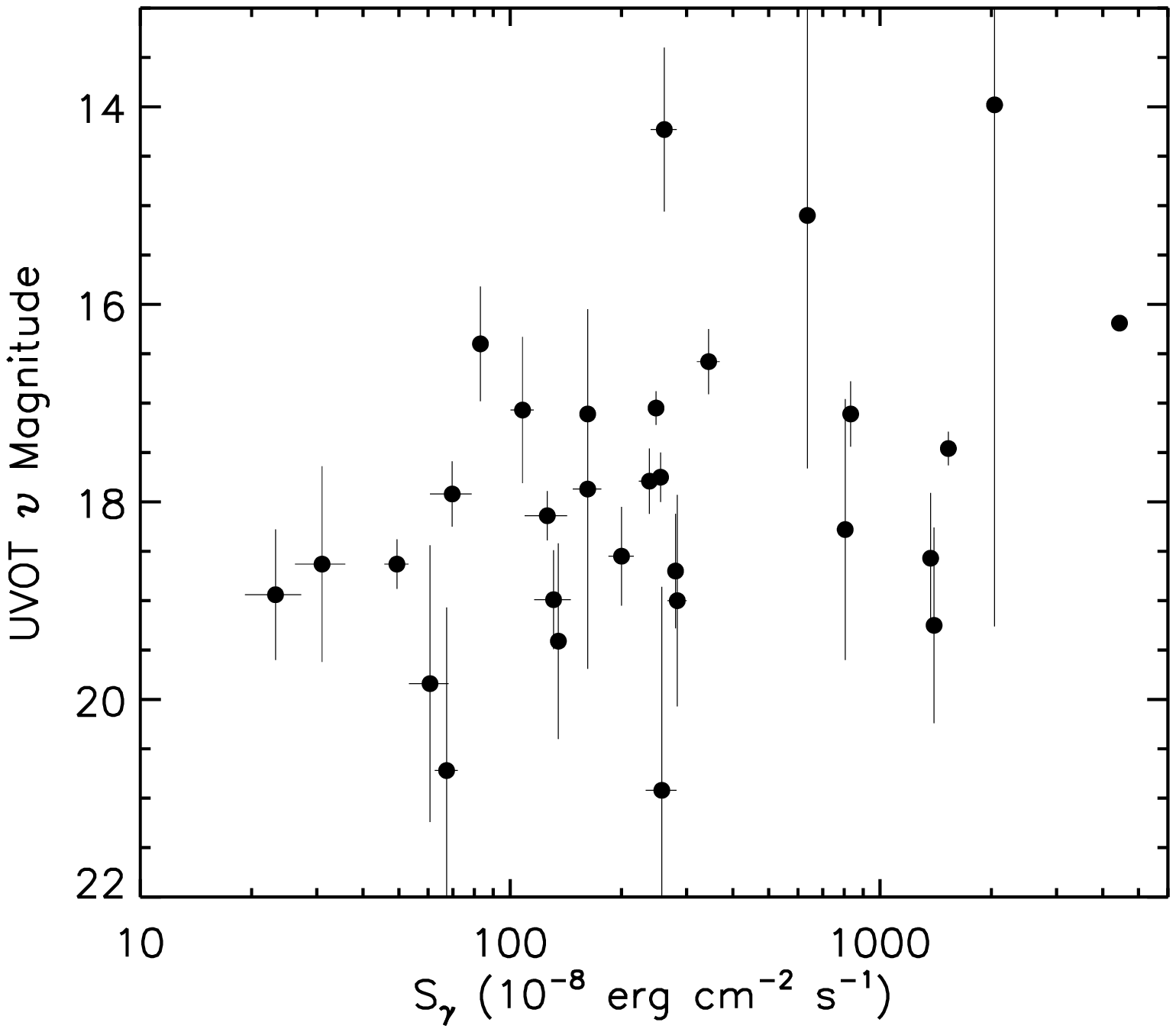}
\caption{Comparison of the prompt BAT fluence 
($S_{\gamma}$) in the $15-150 {\rm ~keV}$ band versus 
the UVOT magnitude in the $v$-band
at $2000 {\rm ~s}$. Using the Spearman rank
correlation, the data are marginally correlated
($p = 0.0184$).}
\label{fig15}   
\end{figure}

\begin{figure}
\epsscale{0.9}
\plotone{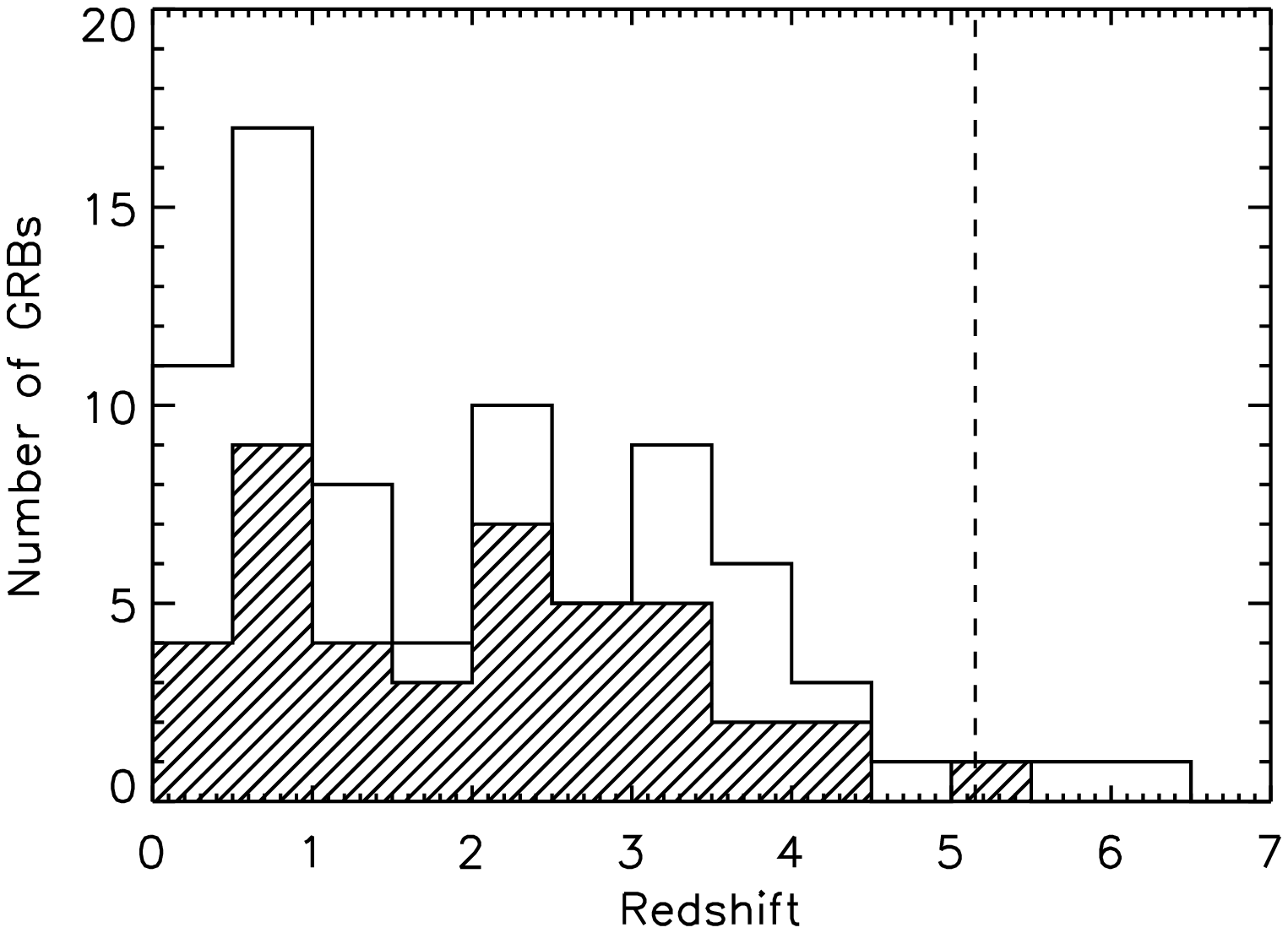}
\caption{Distribution of measured burst redshifts. The 
measurements were made by ground based telescopes and 
not the UVOT. The shaded histogram is for the 
bursts with UVOT detections. The dotted line is the
detection limit for UVOT}
\label{fig-z}   
\end{figure}


\begin{thebibliography}{}

\bibitem[Arimoto {\etal}(2005)]{AM05}
	Arimmoto, M., {\etal} 2005, GRB Coordinates Network, Circular Service, 4550, 1

\bibitem[Barthelmy {\etal}(1995)]{BS1995}
	Barthelmy, S.~D., Butterworth, P., Cline, T.~L., Gehrels, N.,
	Fishman, G.~J., Kouveliotou, C., \& Meegan, C.~A. 1995, 
	\apss, 231, 235

\bibitem[Barthelmy {\etal}(1998)]{BS1998}
	Barthelmy, S.~D., Butterworth, P., Cline, T.~L., \& Gehrels, 
	N., 1998, in AIP Conf. Proc. 428, 4th Huntsville Symp. on
	Gamma-Ray Bursts, ed. C.~A. Meegan, R.~D. Preece, \& T.~M. 
	Koshut (New York: AIP), 139 

\bibitem[Barthelmy {\etal}(2005)]{BS2005}
	Barthelmy, S.~D., {\etal} 2005, \ssr, 120, 143

\bibitem[Blustin {\etal}(2006)]{BAJ06} 
	Blustin, A.~J., {\etal} 2006, \apj, 637, 901

\bibitem[Burrows {\etal}(2005)]{BDN2005}
	Burrows, D.~N., {\etal} 2005, \ssr, 120, 165

\bibitem[Burrows {\etal}(2008)]{BDN08}
	Burrows, D.~N., {\etal} 2008, \apj, in prep

\bibitem[Butler(2007)]{BNR07}
	Butler, N.~R. 2007, \aj, 133, 1027

\bibitem[Campana {\etal}(2006)]{CS06} 
	Campana, S., {\etal} 2006, \nat, 442, 1008

\bibitem[Cardelli {\etal}(1989)]{CCM89} 
	Cardelli, J.~A., Clayton, G.~C., \& Mathis, J.~S. 1989, \apj, 345, 245

\bibitem[De Pasquale {\etal}(2003)]{DM03}
	De Pasquale, M., {\etal} 2003, \apj, 592, 1018

\bibitem[De Pasquale {\etal}(2007)]{DM07}
	De Pasquale, M., {\etal} 2007, \mnras, 377, 1638

\bibitem[Fynbo {\etal}(2001)]{FJU01}
	Fynbo, J.~U., {\etal} 2001, \aap, 369, 373

\bibitem[Gehrels {\etal}(2004)]{GN2004}
	Gehrels, N., {\etal} 2004, \apj, 611, 1005

\bibitem[Goad {\etal}(2007a)]{GM07}
	Goad, M., {\etal} 2007, \aap, 468, 103

\bibitem[Goad {\etal}(2007b)]{GMR07}
	Goad, M., {\etal} 2007, \aap, 476, 1401

\bibitem[Groot {\etal}(1998)]{GPJ98}
	Groot, P.~J., {\etal} 1998, \apj, 493, L27

\bibitem[Grupe {\etal}(2006)]{GD06}
	Grupe, D., {\etal} 2006, \apj, 645, 464

\bibitem[Halpern {\etal}(1998)]{HJP98} 
	Halpern, J.~P., Thorstensen, J.~R., Helfand, D.~J., 
	\& Costa, E. 1998, \nat, 393, 41

\bibitem[Hurley {\etal}(2005a)]{HK05}
	Hurley, K., {\etal} 2005, \apjs, 156, 217

\bibitem[Hurley {\etal}(2005b)]{HK05b}
	Hurley, K., {\etal} 2005, GRB Coordinates Network, Circular Service, 4172, 1

\bibitem[Jakobsson {\etal}(2004)]{JP04}
	Jakobsson, P., {\etal} 2004, \apj, 617, L21

\bibitem[Jester {\etal}(2005)]{JS05}
	Jester, S., {\etal} 2005, \aj, 130, 873

\bibitem[Kalberla {\etal}(2005)]{KPMW05}
	Kalberla, P.~M.~W., Burton, W.~B., Hartmann, D., Arnal, E.~M.,
	Bajaja, E.,	Morras, R., \& P\"oppel, W.~G.~L. 2005, \aap, 440, 775

\bibitem[Kann {\etal}(2007)]{KDA07}
	Kann, D.~A., {\etal} 2007, \apj, submitted (arXiv:0712.2186)

\bibitem[Kawai {\etal}(2005)]{KN05}
	Kawai, N., {\etal} 2005, GRB Coordinates Network, Circular Service, 4359, 1

\bibitem[Kuin \& Rosen(2008)]{KR08}
	Kuin, N.~P.~M., \& Rosen, S.~R. 2008, \mnras, 383, 383

\bibitem[Mason {\etal}(2006)]{MKO06}
	Mason, K.~O., {\etal} 2006, \apj, 639, 311

\bibitem[Meegan {\etal}(1992)]{MCA92} 
	Meegan, C.~A., Fishman, G.~J., Wilson, R.~B., Horack, J.~M.,
	Brock, M.~N., Paciesas, W.~S., Pendleton, G.~N., \&
	Kouveliotou, C. 1992, \nat, 355, 143

\bibitem[Monet {\etal}(2003)]{MDG03}
	Monet, D.~G., {\etal} 2003, \aj, 125, 984

\bibitem[Morgan {\etal}(2008)]{MA08}
	Morgan, A.~N., {\etal} 2008, \apj, accepted

\bibitem[Oates {\etal}(2007)]{OS2007}
	Oates, S.~R., {\etal} 2007, \mnras, 380, 270

\bibitem[Oates {\etal}(2008)]{OS08}
	Oates, S.~R., {\etal} 2008, \mnras, in prep

\bibitem[Olive {\etal}(2005)]{OJ05}
	Olive, J.~-F., {\etal} 2005, GRB Coordinates Network, Circular Service, 4124, 1

\bibitem[Perri {\etal}(2007)]{PM07}
	Perri, M., {\etal} 2007, \aap, 471, 83

\bibitem[Poole {\etal}(2008)]{PTS2007}
	Poole, T.~S., {\etal} 2008, \mnras, 383, 627

\bibitem[Ricker (1997)]{RG1997}
	Ricker, G.~R. 1997, in All-Sky X-Ray Observations in the 
	Next Decade, ed. M. Matsuoka \& N. Kawai (Japan: RIKEN), 366 

\bibitem[Rol {\etal}(2005)]{RE05}
	Rol, E., Wijers, R.~A.~M.~J., Kouveliotou, C., Kaper, L.,
	\& Kaneko, Y. 2005, \apj, 624, 868

\bibitem[Roming {\etal}(2005)]{RPWA2005}
	Roming, P.~W.~A., {\etal} 2005, \ssr, 120, 95

\bibitem[Roming {\etal}(2006a)]{RPWA06}
	Roming, P.~W.~A., {\etal} 2006, \apj, 651, 985

\bibitem[Roming {\etal}(2006b)]{RPWA06b}
	Roming, P.~W.~A., {\etal} 2006, \apj, 652, 1416

\bibitem[Sakamoto {\etal}(2005)]{ST05}
	Sakamoto, T., {\etal} 2005, GRB Coordinates Network, Circular Service, 3189, 1

\bibitem[Sakamoto {\etal}(2007)]{ST2007}
	Sakamoto, T., {\etal} 2007, \apjs, 175, 179

\bibitem[Schady {\etal}(2007)]{SP07}
	Schady, P., {\etal} 2007, \mnras, 380, 1041

\bibitem[Schlegel {\etal}(1998)]{SFD98}
	Schlegel, D.~J., Finkbeiner, D.~P., \& Davis, M. 1998, \apj, 500, 525

\bibitem[Still {\etal}(2005)]{SM05}
	Still, M., {\etal} 2005, \apj, 635, 1187

\bibitem[York {\etal}(2000)]{YDG00}
	York, D.~G., {\etal} 2000, \aj, 120, 1579

\end{thebibliography}
\end{document}